\documentclass[aps,prd,amsmath,amssymb,twocolumn,preprintnumbers]{revtex4-2}
\usepackage[english]{babel}
\usepackage{bm}
\usepackage{bbm}
\usepackage{mathtools}
\usepackage{graphicx}
\usepackage{tikz}
\usetikzlibrary{arrows,shapes,backgrounds,calc,positioning,tikzmark,patterns,automata,decorations.markings}
\tikzset{fontscale/.style={font=\relsize{#1}}}
\tikzset{->-/.style={decoration={
  markings,
  mark=at position #1 with {\arrow{>}}},postaction={decorate}}}
\tikzset{-<-/.style={decoration={
  markings,
  mark=at position #1 with {\arrow{<}}},postaction={decorate}}}

\tikzset{cross/.style={cross out,draw,minimum size=2*(#1-\pgflinewidth),inner sep=0pt, outer sep=0pt}}

\tikzset{
  pics/carc/.style args={#1:#2:#3}{
    code={
      \draw[pic actions] (0,0) -- (#1:#3) arc(#1:#2:#3) -- cycle;
    }
  }
}
\usepackage{picture}
\usepackage{cancel}
\usepackage{tabularx}
\usepackage{tensor}
\usepackage{float}
\usepackage{placeins}
\usepackage{fancyhdr}
\usepackage{sidecap}
\usepackage{xcolor}
\usepackage{scalerel}
\usepackage{relsize}
\usepackage[absolute,overlay]{textpos}
\usepackage[customcolors]{hf-tikz}
\usepackage{cals}
\usepackage{enumitem}
\setlist{nolistsep}
\usepackage{pifont}
\usepackage[export]{adjustbox} 
\usepackage[most]{tcolorbox}
\usepackage{varwidth}
\tcbuselibrary{skins}
\usepackage[pdfborder={0 0 0}]{hyperref}
\definecolor{darkblue}{rgb}{0,0,.5}
\definecolor{darkgreen}{rgb}{0,0.5,.5}
\definecolor{darkyellow}{rgb}{0.5,0.5,0}
\definecolor{fhl}{rgb}{1,0,0}
\hypersetup{colorlinks=true, breaklinks=true, linkcolor=darkblue, menucolor=darkblue, urlcolor=darkblue, linktocpage=true}

\newtcolorbox{hlbox}[2][red]{
colbacktitle=#1!10,
colback=white!95!#1,
coltitle=black,
fonttitle=\bfseries,
colframe=#1!50,
boxrule=0.5pt,
titlerule=0pt,
title={\strut#2},
arc=3pt,
middle=0pt,
boxsep=0pt,
left skip=0pt,
right skip=0pt}

\let\originalleft\left
\let\originalright\right
\renewcommand{\left}{\mathopen{}\mathclose\bgroup\originalleft}
\renewcommand{\right}{\aftergroup\egroup\originalright}
\newcommand{\e}{\operatorname{e}}
\newcommand{\SU}[1]{\operatorname{SU}\left(#1\right)}

\newcommand{\Un}[1]{\operatorname{U}\left(#1\right)}

\newcommand{\su}[1]{\mathfrak{su}\left(#1\right)}

\newcommand{\of}[1]{\left(#1\right)}

\newcommand{\sof}[1]{\bigl(\big.#1\big.\bigr)}
\newcommand{\ssof}[1]{(#1)}
\newcommand{\fof}[1]{\left[\right.#1\left.\right]}

\newcommand{\sfof}[1]{\bigl[\big.#1\big.\bigr]}
\newcommand{\cof}[1]{\left\{\right.#1\left.\right\}}

\newcommand{\scof}[1]{\bigl\{\big.#1\big.\bigr\}}
\newcommand{\loint}[1]{\left(#1\right]}

\newcommand{\trace}{\operatorname{tr}}

\newcommand{\avof}[1]{\left\langle #1\right\rangle}

\newcommand{\repart}{\operatorname{Re}}

\newcommand{\ii}{\mathrm{i}}
\newcommand{\idd}[2]{\mathrm{d}^{#2}#1}
\newcommand{\dd}{\mathrm{d}}

\newcommand{\order}[1]{\mathcal{O}\big(#1\big)}

\newcommand{\id}{\mathbbm{1}}

\newcommand{\abs}[1]{\left| #1\right|}
\newcommand{\ssabs}[1]{| #1|}

\renewcommand*\[{\begin{equation}}
\renewcommand*\]{\end{equation}}
\renewcommand*\hat[1]{\widehat{#1}}
\let\oldstackrel\stackrel
\renewcommand*\stackrel[2]{{\scriptstyle\oldstackrel{#1}{#2}}}

\definecolor{emphcol}{rgb}{1.,0,0}
\let\oldemph\emph
\renewcommand*\emph[1]{\oldemph{\textcolor{emphcol}{#1}}}

\newcommand{\ucases}[1]{\begin{cases}#1\end{cases}}

\makeatletter
\pgfarrowsdeclare{open cap}{open cap}
{\pgfarrowsleftextend{+0pt}\pgfarrowsrightextend{+0.5\pgflinewidth}}
{
  \pgfmathsetlength{\pgfutil@tempdimb}{.5*\pgflinewidth-.5*\pgfinnerlinewidth}%
  \pgfsetlinewidth{\pgfutil@tempdimb}
  \pgfsetbuttcap
  \pgfsetdash{}{0pt}
  \pgfmathsetlength{\pgfutil@tempdima}{.5*\pgfutil@tempdimb+.5*\pgfinnerlinewidth}%
  \pgfpathmoveto{\pgfqpoint{0pt}{\pgfutil@tempdima}}
  \pgfpatharc{270}{90}{-\pgfutil@tempdima}
  \pgfusepathqstroke
}
\makeatother
\allowdisplaybreaks

\newcommand{\changed}[1]{#1}
\newcommand{\changedn}[1]{#1}

\begin{document}
\title{Bulk-preventing actions for SU(N) gauge theories}

\author{Tobias Rindlisbacher}
\email{trindlis@itp.unibe.ch}
\affiliation{Albert Einstein Center for Fundamental Physics \& Institute for Theoretical Physics, University of Bern, Sidlerstrasse 5, CH-3012 Bern, Switzerland}
\author{Kari Rummukainen}
\email{kari.rummukainen@helsinki.fi}
\author{Ahmed Salami}
\email{ahmed.salami@helsinki.fi}
\affiliation{Department of Physics \& Helsinki Institute of Physics, P.O. Box 64, FI-00014 University of Helsinki, Finland}

\begin{abstract}
Lattice gauge field theories may suffer from unphysical "bulk" phase transitions at strong lattice gauge coupling.
We introduce a one-parameter family of lattice $\SU{N}$ gauge actions which, when used in combination with an HMC update algorithm, prevents the appearance of the bulk phase transition. We briefly discuss the (presumed) mechanism behind the prevention of the bulk transition and present test results for different $\SU{N}$ gauge groups.
\end{abstract}
\maketitle

\section{Introduction}

In asymptotically free gauge theories on the lattice the continuum limit is obtained when the bare lattice gauge coupling vanishes.  In practice lattice simulations are always done at finite lattice spacing, and as long as the coupling constant is sufficiently small we can analytically extrapolate the results to the continuum limit.  The range of lattice spacings (or coupling constants) is often limited by the emergence of an unphysical ``bulk phase'' at strong lattice gauge coupling, which prevents the analytical connection to the continuum phase from this region. The value of the lattice coupling where the transition to the bulk phase happens depends on the lattice gauge group, the choice of the gauge action and the matter content.

The problem of the bulk phase transition becomes acute in $\SU{N}$ gauge theories with large number of colors and in models with large number of fermion degrees of freedom. In pure gauge $\SU{N}$ theories with the standard Wilson plaquette action the bulk transition is a rapid cross-over if $N\le 4$ but becomes an increasingly strong first order transition for $N \ge 5$ \cite{Lucini:2005vg}. Adding fermionic degrees of freedom slows down the evolution of the coupling constant (i.e. the magnitude of the $\beta$-function is smaller) and also increase the effective lattice gauge coupling \cite{Hasenfratz:1993az}. Depending on the physical case of interest, these effects require one to use strong bare coupling.
This happens especially in infrared (near-)conformal models, where the coupling runs very slowly, for example in $\SU{2}$ with large numbers of fundamental fermions \cite{Bursa:2010xn,Karavirta:2011zg,Leino:2019qwk,Rantaharju:2021iro,Rindlisbacher:2021hhh}, $\SU{2}$ with adjoint fermions \cite{Hietanen:2008mr,Bursa:2009we, Rantaharju:2015cne}, $\SU{3}$ with large Nf \cite{LSD:2014nmn,Hasenfratz:2022zsa,LatticeStrongDynamics:2018hun}, and $\SU{4}$ with fermions in the antisymmetric representation~\cite{DeGrand:2014cea}. 

\changed{In this work we} present a local lattice \changed{gauge action, defined on elementary plaquettes}, which efficiently removes the transition to the bulk phase. Our approach is related to the "dislocation prevention" method by DeGrand {\em et al.}~\cite{DeGrand:2014rwa}. For $\Un{1}$ and $\SU{2}$ gauge groups also the topological gauge actions that restict the plaquette magnitude~\cite{Bornyakov:1991gq,Akerlund:2015zha,Nogradi:2018ivi} are to some extent related. \changed{The latter stops being} the case for $\SU{N}$ gauge groups with \changed{$N>2$, as the plaquette eigenvalues are then no longer determined by the trace of the plaquette.} We note that our approach removes merely the bulk phase but does not restrict in any way gauge-topology fluctuations.

Following Wilson's prescription, the lattice discretization of a $\SU{N}$ gauge theory is obtained by promoting the Lie algebra valued continuum gauge field, 
\[
A_{\mu}\of{x'}=\sum_{a} A^{a}_{\mu}\of{x'}\,T^{a}\,\in\,\su{N}\ ,\label{eq:contgaugefield}
\]
with $\cof{T^{a}}_{a=1,\ldots,N^2-1}$ being a basis of $\su{N}$, to Lie group valued link variables,
\[
U_{\mu}\of{x}\,=\,\mathcal{P}\e^{\ii\,\int_{a x}^{a\,(x+\hat{\mu})}\dd{x'}\,A_{\mu}\ssof{x'}}\,\in\,\SU{N}\  ,\label{eq:latgaugefield}
\]
which can be interpreted as the gauge-parallel transporters along the link between a site $x$ and a neighboring site $x+\hat{\mu}$. The leading $\mathcal{P}$ on the right-hand side of \eqref{eq:latgaugefield} indicate that path-ordering should be applied when evaluating the exponential of the line-integral. The relation between the coordinate $x'\in\mathbb{R}^4$ in \eqref{eq:contgaugefield} and the coordinate $x\in\mathbb{Z}^4$ in \eqref{eq:latgaugefield} is given by $x'=a\,x$, where $a$ is the lattice spacing, and $\hat{\mu}$ refers to the unit-vector in $\mu$-direction. Parallel transporters over longer distances are then expressed as products of consecutive link variables and a lattice gauge action can be defined in terms of link variables by requiring that in the limit $\of{a\to 0}$ the lattice gauge action converges to the continuum gauge action,
\[
S_{G}=\frac{1}{\changed{2}\,g_0^2}\int\idd{x'}{4}\trace\of{F_{\mu\nu}\of{x'}F_{\mu\nu}\of{x'}}\ .
\]
Wilson proposed the gauge action~\cite{Wilson:1974sk}
\[
S_{G,W}=\frac{\beta}{N}\sum\limits_{x}\sum\limits_{\mu<\nu}\repart\trace\sof{\id-U_{\mu\nu}\of{x}}\ ,\label{eq:wilsongaugeaction}
\]
which, as is well known, satisfies the above condition and is here
written in terms of the inverse bare gauge coupling $\beta=2\,N/g_0^2$ and the plaquette variables
\[
U_{\mu\nu}\of{x}=U_{\mu}\of{x}\,U_{\nu}\of{x+\hat{\mu}}\,U^{\dagger}_{\mu}\of{x+\hat{\nu}}\,U^{\dagger}_{\nu}\of{x}\ .\label{eq:plaquettevariable}
\]

The gauge action \eqref{eq:wilsongaugeaction} and improved versions of it~\cite{Luscher:1984xn} are the most commonly used in Monte Carlo studies of $\SU{N}$ lattice gauge theories. They are, however, not unique and might not be the best choice for the study of lattice gauge theories at strong coupling, as they allow the gauge system to enter the bulk phase. \changed{Despite its name, the bulk phase} is not necessarily a proper phase, but simply a region in parameter space of the lattice theory where lattice artifacts dominate in ensemble averages. As a consequence the relation between lattice and continuum results becomes very complicated or can even be lost completely if bulk and continuum phase are separated by a first order transition.

\section{Avoiding the lattice bulk phase}\label{sec:avoidinglatbulkphase}

In this section we propose a characterization of "bulk configurations" in $\SU{N}$ lattice gauge systems, which allows for the definition of a family of lattice gauge actions, that separate such bulk configurations from regular ones by an infinite potential barrier, while still yielding the same naive continuum limit as Wilson's plaquette gauge action. In combination with a hybrid Monte Carlo (HMC) update algorithm, the new gauge actions prevent the gauge system from entering a bulk phase.

\subsection{Motivation in $\Un{1}$}\label{ssec:motivationinu1}
\begin{figure}[htb]
\centering
\includegraphics[width=0.85\linewidth,keepaspectratio]{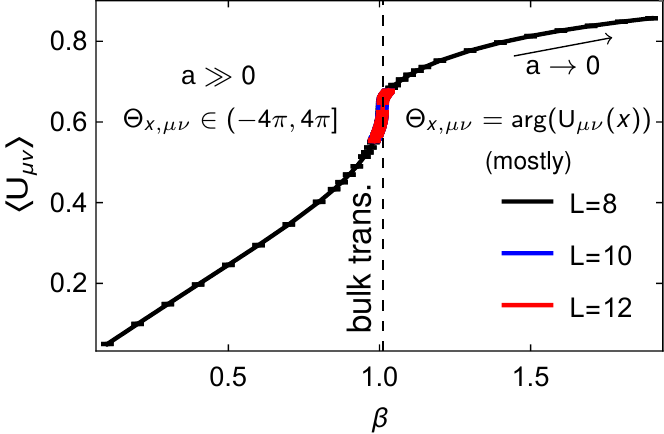}
\caption{The $\Un{1}$ lattice gauge theory with Wilson gauge action \eqref{eq:wilsongaugeactionu1} undergoes a bulk-transition at $\beta=\beta_b\approx 1$. For $\beta<\beta_b$, the system is in the bulk phase, where the lattice spacing, $a$, can be considered large and $\Theta_{x,\mu\nu}$ from \eqref{eq:totplaqphaseu1} explores the full $\loint{-4\pi,4\pi}$-interval. For $\beta>\beta_b$, the system is in the continuum phase, where the lattice spacing tends to zero if $\beta\to\infty$. In this phase, $\Theta_{x,\mu\nu}$ can still be outside the $\loint{-\pi,\pi}$-interval, but the fraction of such plaquettes quickly drops as $\beta$ is increased and one mostly has $\Theta_{x,\mu\nu}=\arg\of{U_{\mu\nu}\of{x}}$ . }
\label{fig:phasediagramu1}
\end{figure}

In the $\Un{1}$ case, the Wilson gauge action in \eqref{eq:wilsongaugeaction} reduces to
\[
S_{G,W}=\beta\,\sum\limits_{x}\sum\limits_{\mu<\nu}\repart\sof{1-U_{\mu\nu}\of{x}}\ ,\label{eq:wilsongaugeactionu1}
\]
and the Abelian link variables can be written as
\[
U_{\mu}\of{x}=\e^{\ii\,\theta_{x,\mu}}\quad\text{with}\quad \theta_{x,\mu}=a\,A_{\mu}\of{x}\,\in\,\loint{-\pi,\pi}\ .\label{eq:abelianlinkvar}
\]
Let us now define,
\[
\Theta_{x,\mu\nu}=\theta_{x,\mu}+\theta_{x+\hat{\mu},\nu}-\theta_{x+\hat{\nu},\mu}-\theta_{x,\nu}\,\in\,\loint{-4\pi,4\pi}\ ,\label{eq:totplaqphaseu1}
\]
and note that while $\Theta_{x,\mu\nu}$ in \eqref{eq:totplaqphaseu1} can vary in the interval $\loint{-4\pi,4\pi}$, the gauge action \eqref{eq:wilsongaugeactionu1} depends only on
\[
\arg\of{U_{\mu\nu}\of{x}}\,\in\,\loint{-\pi,\pi}\ .\label{eq:argplaqu1}
\]
As illustrated in Fig.~\ref{fig:phasediagramu1}, the gauge action \eqref{eq:wilsongaugeactionu1} produces a bulk-transition at $\beta=\beta_b\approx 1$. For $\beta<\beta_b$, the system is in the bulk phase, where the lattice spacing, $a$, can be considered large and $\Theta_{x,\mu\nu}$ from \eqref{eq:totplaqphaseu1} explores the full $\loint{-4\pi,4\pi}$-interval. For $\beta>\beta_b$, the system is in the continuum phase, where the lattice spacing tends to zero if $\of{\beta\to\infty}$. In this phase, $\Theta_{x,\mu\nu}$ can still be outside the $\loint{-\pi,\pi}$-interval, but the fraction of such plaquettes quickly drops as $\of{\beta\to\infty}$ and most of the time, one has that $\Theta_{x,\mu\nu}=\arg\of{U_{\mu\nu}\of{x}}$ . 

\begin{figure}[htb]
\centering
\begin{enumerate}[label=(\alph*)]
\item\label{en:linkwrap} link wraps around $\loint{-\pi,\pi}$-interval:\\[-17pt]
\begin{center}
\begin{tikzpicture}[scale=0.27,nodes={inner sep=0},every node/.style={transform shape}]
  \def\tha{-155.0}
  \pgfmathparse{\tha<0 ? "blue" : "red"}
  \edef\cola{\pgfmathresult}
  \def\thb{-60.0}
  \pgfmathparse{\thb<0 ? "blue" : "red"}
  \edef\colb{\pgfmathresult}
  \def\thc{135.0}
  \pgfmathparse{\thc<0 ? "blue" : "red"}
  \edef\colc{\pgfmathresult}
  \def\thd{100.0}
  \pgfmathparse{\thd<0 ? "blue" : "red"}
  \edef\cold{\pgfmathresult}

  \pgfmathparse{\tha+\thb+\thc+\thd}
  \edef\thtot{\pgfmathresult}
  
  \pgfmathparse{\thtot<0 ? "blue" : "red"}
  \edef\coltot{\pgfmathresult}
  
  \def\thaalt{155.0}
  \pgfmathparse{\thaalt<0 ? "blue" : "red"}
  \edef\colaalt{\pgfmathresult}
  
  \pgfmathparse{\thaalt+\thb+\thc+\thd}
  \edef\thtotalt{\pgfmathresult}

  \pgfmathparse{\thtotalt<0 ? "blue" : "red"}
  \edef\coltotalt{\pgfmathresult}

  \pgfpointtransformed{\pgfpointxy{1}{1}};
  \pgfgetlastxy{\vx}{\vy};
  \begin{scope}[node distance=\vx and \vy,fontscale=-1]
    \draw[-<-=.18,thick,black] (0,0) -- (4,0);
    \draw[-<-=.18,thick,black] (4,0) -- (4,4);
    \draw[-<-=.18,thick,black] (4,4) -- (0,4);
    \draw[-<-=.18,thick,black] (0,4) -- (0,0);
    
    \draw node[black,below,scale=3.5] at (0,-0.2) {$x\phantom{\hat{\mu}}$};
    \draw node[black,below,scale=3.5] at (4.2,-0.2) {$x+\hat{\mu}$};
    \draw node[black,above,scale=3.5] at (5.1,4.2) {$x+\hat{\mu}+\hat{\nu}$};
    \draw node[black,above,scale=3.5] at (-0.2,4.2) {$x+\hat{\nu}\vphantom{\hat{\mu}}$};
    \draw[thick,fill=white] (2,0) circle (0.8) pic[\cola,fill=\cola!50!white]{carc=90:90-\tha:0.8};  
    \draw[thick,fill=white] (4,2) circle (0.8) pic[\colb,fill=\colb!50!white]{carc=90:90-\thb:0.8};
    \draw[thick,fill=white] (2,4) circle (0.8) pic[\colc,fill=\colc!50!white]{carc=90:90-\thc:0.8};    
    \draw[thick,fill=white] (0,2) circle (0.8) pic[\cold,fill=\cold!50!white]{carc=90:90-\thd:0.8};
    \draw[thick,fill=white] (2,1.9) circle (0.8) pic[\coltot,fill=\coltot!50!white,->]{carc=90:90-\thtot:0.8};
    \draw[black,fill=white] (0,0) circle (5pt); 
    \draw[black,fill=white] (4,0) circle (5pt);
    \draw[black,fill=white] (4,4) circle (5pt);
    \draw[black,fill=white] (0,4) circle (5pt);

    \def\xdisp{10}
    \draw node[black,scale=4.5] at (2+0.5*\xdisp,2.0) {$\Longrightarrow$};    
    
    \draw[-<-=.18,thick,black] (\xdisp,0) -- (4+\xdisp,0);
    \draw[-<-=.18,thick,black] (4+\xdisp,0) -- (4+\xdisp,4);
    \draw[-<-=.18,thick,black] (4+\xdisp,4) -- (\xdisp,4);
    \draw[-<-=.18,thick,black] (\xdisp,4) -- (\xdisp,0);
    
    \draw node[black,below,scale=3.5] at (\xdisp,-0.2) {$x\phantom{\hat{\mu}}$};
    \draw node[black,below,scale=3.5] at (4.2+\xdisp,-0.2) {$x+\hat{\mu}$};
    \draw node[black,above,scale=3.5] at (5.1+\xdisp,4.2) {$x+\hat{\mu}+\hat{\nu}$};
    \draw node[black,above,scale=3.5] at (\xdisp-0.2,4.2) {$x+\hat{\nu}\vphantom{\hat{\mu}}$};
    \draw[thick,fill=white] (2+\xdisp,0) circle (0.8) pic[\colaalt,fill=\colaalt!50!white]{carc=90:90-\thaalt:0.8};  
    \draw[thick,fill=white] (4+\xdisp,2) circle (0.8) pic[\colb,fill=\colb!50!white]{carc=90:90-\thb:0.8};
    \draw[thick,fill=white] (2+\xdisp,4) circle (0.8) pic[\colc,fill=\colc!50!white]{carc=90:90-\thc:0.8};    
    \draw[thick,fill=white] (\xdisp,2) circle (0.8) pic[\cold,fill=\cold!50!white]{carc=90:90-\thd:0.8};
    \draw[thick,fill=white] (2+\xdisp,1.9) circle (0.8) pic[\coltotalt,fill=\coltotalt!50!white,->]{carc=90:90-\thtotalt:0.8};
    \draw[black,fill=white] (\xdisp,0) circle (5pt); 
    \draw[black,fill=white] (4+\xdisp,0) circle (5pt);
    \draw[black,fill=white] (4+\xdisp,4) circle (5pt);
    \draw[black,fill=white] (\xdisp,4) circle (5pt); 
  \end{scope}
\end{tikzpicture}\\[-20pt]
\end{center}
\item\label{en:nolinkwrap} $\Theta_{x,\mu\nu}$ grows continuously:\\[-17pt]
\begin{center}
\begin{tikzpicture}[scale=0.27,nodes={inner sep=0},every node/.style={transform shape}]
  \def\tha{20.0}
  \pgfmathparse{\tha<0 ? "blue" : "red"}
  \edef\cola{\pgfmathresult}
  \def\thb{40.0}
  \pgfmathparse{\thb<0 ? "blue" : "red"}
  \edef\colb{\pgfmathresult}
  \def\thc{50.0}
  \pgfmathparse{\thc<0 ? "blue" : "red"}
  \edef\colc{\pgfmathresult}
  \def\thd{60.0}
  \pgfmathparse{\thd<0 ? "blue" : "red"}
  \edef\cold{\pgfmathresult}

  \pgfmathparse{\tha+\thb+\thc+\thd}
  \edef\thtot{\pgfmathresult}
  
  \pgfmathparse{\thtot<0 ? "blue" : "red"}
  \edef\coltot{\pgfmathresult}
  
  \def\thaalt{40.0}
  \pgfmathparse{\thaalt<0 ? "blue" : "red"}
  \edef\colaalt{\pgfmathresult}
  
  \pgfmathparse{\thaalt+\thb+\thc+\thd}
  \edef\thtotalt{\pgfmathresult}
  
  \pgfmathparse{\thtotalt<0 ? "blue" : "red"}
  \edef\coltotalt{\pgfmathresult}

  \pgfpointtransformed{\pgfpointxy{1}{1}};
  \pgfgetlastxy{\vx}{\vy};
  \begin{scope}[node distance=\vx and \vy,fontscale=-1]
    \draw[-<-=.18,thick,black] (0,0) -- (4,0);
    \draw[-<-=.18,thick,black] (4,0) -- (4,4);
    \draw[-<-=.18,thick,black] (4,4) -- (0,4);
    \draw[-<-=.18,thick,black] (0,4) -- (0,0);
    
    \draw node[black,below,scale=3.5] at (0,-0.2) {$x\phantom{\hat{\mu}}$};
    \draw node[black,below,scale=3.5] at (4.2,-0.2) {$x+\hat{\mu}$};
    \draw node[black,above,scale=3.5] at (5.1,4.2) {$x+\hat{\mu}+\hat{\nu}$};
    \draw node[black,above,scale=3.5] at (-0.2,4.2) {$x+\hat{\nu}\vphantom{\hat{\mu}}$};
    \draw[thick,fill=white] (2,0) circle (0.8) pic[\cola,fill=\cola!50!white]{carc=90:90-\tha:0.8};  
    \draw[thick,fill=white] (4,2) circle (0.8) pic[\colb,fill=\colb!50!white]{carc=90:90-\thb:0.8};
    \draw[thick,fill=white] (2,4) circle (0.8) pic[\colc,fill=\colc!50!white]{carc=90:90-\thc:0.8};    
    \draw[thick,fill=white] (0,2) circle (0.8) pic[\cold,fill=\cold!50!white]{carc=90:90-\thd:0.8};
    \draw[thick,fill=white] (2,1.9) circle (0.8) pic[\coltot,fill=\coltot!50!white,->]{carc=90:90-\thtot:0.8};
    \draw[black,fill=white] (0,0) circle (5pt); 
    \draw[black,fill=white] (4,0) circle (5pt);
    \draw[black,fill=white] (4,4) circle (5pt);
    \draw[black,fill=white] (0,4) circle (5pt);

    \def\xdisp{10}
    \draw node[black,scale=4.5] at (2+0.5*\xdisp,2.0) {$\Longrightarrow$};    
    
    \draw[-<-=.18,thick,black] (\xdisp,0) -- (4+\xdisp,0);
    \draw[-<-=.18,thick,black] (4+\xdisp,0) -- (4+\xdisp,4);
    \draw[-<-=.18,thick,black] (4+\xdisp,4) -- (\xdisp,4);
    \draw[-<-=.18,thick,black] (\xdisp,4) -- (\xdisp,0);
    
    \draw node[black,below,scale=3.5] at (\xdisp,-0.2) {$x\phantom{\hat{\mu}}$};
    \draw node[black,below,scale=3.5] at (4.2+\xdisp,-0.2) {$x+\hat{\mu}$};
    \draw node[black,above,scale=3.5] at (5.1+\xdisp,4.2) {$x+\hat{\mu}+\hat{\nu}$};
    \draw node[black,above,scale=3.5] at (\xdisp-0.2,4.2) {$x+\hat{\nu}\vphantom{\hat{\mu}}$};
    \draw[thick,fill=white] (2+\xdisp,0) circle (0.8) pic[\colaalt,fill=\colaalt!50!white]{carc=90:90-\thaalt:0.8};  
    \draw[thick,fill=white] (4+\xdisp,2) circle (0.8) pic[\colb,fill=\colb!50!white]{carc=90:90-\thb:0.8};
    \draw[thick,fill=white] (2+\xdisp,4) circle (0.8) pic[\colc,fill=\colc!50!white]{carc=90:90-\thc:0.8};    
    \draw[thick,fill=white] (\xdisp,2) circle (0.8) pic[\cold,fill=\cold!50!white]{carc=90:90-\thd:0.8};
    \draw[thick,fill=white] (2+\xdisp,1.9) circle (0.8) pic[\coltotalt,fill=\coltotalt!50!white,->]{carc=90:90-\thtotalt:0.8};
    \draw[black,fill=white] (\xdisp,0) circle (5pt); 
    \draw[black,fill=white] (4+\xdisp,0) circle (5pt);
    \draw[black,fill=white] (4+\xdisp,4) circle (5pt);
    \draw[black,fill=white] (\xdisp,4) circle (5pt); 
  \end{scope}
\end{tikzpicture}
\end{center}
\end{enumerate}
\caption{The figure illustrates the two different ways in which $\Theta_{x,\mu\nu}$ form \eqref{eq:totplaqphaseu1} can grow beyond the $\loint{-\pi,\pi}$-interval: a) a single link (between $x$ and $x+\hat{\mu}$) winds around $\loint{\pi,\pi}$ which adds almost $2\,\pi$ to $\Theta_{x,\mu\nu}$; b) no link wraps around the $\loint{-\pi,\pi}$-interval, $\Theta_{x,\mu\nu}$ grows continuously bigger than $\pi$.}
\label{fig:contvsdiscontwindingu1}
\end{figure}

The fact that plaquettes with $\Theta_{x,\mu\nu}\notin\loint{-\pi,\pi}$ also appear in the continuum phase indicates that the value of $\Theta_{x,\mu\nu}$ by itself cannot be used to distinguish bulk from continuum configurations. 
However, as illustrated in Fig.~\ref{fig:contvsdiscontwindingu1}, \changed{for smoothly varying link variables, $U_{\mu}\of{x}$, one can identify} two qualitatively different ways in which plaquettes with $\Theta_{x,\mu\nu}\notin\loint{-\pi,\pi}$ can be produced from a configuration in which initially $\Theta_{x,\mu\nu}=\arg\of{U_{\mu\nu}\of{x}}$ is satisfied for all plaquettes: 
\begin{enumerate}[label=(\alph*)]
\item\label{en:caselinkwrap} the phase of one of the links of a given plaquette can move across the boundary of the $\loint{-\pi,\pi}$-interval and wrap around, which adds roughly $\pm 2\,\pi$ to the plaquette's phase angle $\Theta_{x,\mu\nu}$. As indicated in part~\ref{en:linkwrap} of Fig.~\ref{fig:contvsdiscontwindingu1}, this kind of plaquette wrapping can happen also if $\arg\of{U_{\mu\nu}\of{x}}$ is close to $0$. We note that the wrapping link will produce a shift of roughly $\pm 2\,\pi$ in the phase angle $\Theta_{x,\mu\nu}$ of all plaquettes that contain this link and not just in the phase angle of the plaquette under consideration;
\item\label{en:casenolinkwrap} none of the links of a given plaquette wraps around the $\loint{-\pi,\pi}$-interval. Instead, as indicated in part~\ref{en:nolinkwrap} of Fig.~\ref{fig:contvsdiscontwindingu1}, the plaquette's phase angle $\Theta_{x,\mu\nu}$ has grown close to the $\loint{-\pi,\pi}$-interval boundary and finally crosses it continuously as one of the link variables undergoes a change that causes $\Theta_{x,\mu\nu}$ to grow further. Unlike in case~\ref{en:caselinkwrap}, this continuous plaquette wrapping can take place for single plaquettes. This can produce metastabilities as wrapped and unwrapped plaquettes pull in opposite directions on shared links when the action is minimized. In order to approach the continuum limit, the wrapping needs to be undone.
\end{enumerate}
As case~\ref{en:casenolinkwrap} can occur only if individual plaquette angles are allowed to grow close to $\pm\pi$, this case is likely to occur only in the bulk phase, where $\beta$ is so small that the plaquette action~\eqref{eq:wilsongaugeactionu1} cannot grow sufficiently large as to oppose the entropy-driven randomization of the plaquette variables in the system. We will therefore introduce in Sec.~\ref{ssec:bulkprevaction} a family of actions that will prevent plaquette wrappings of type~\ref{en:casenolinkwrap}. As it turns out, this is sufficient to get rid of the bulk transition. First, however, we discuss in Sec.~\ref{ssec:situationinsun} how the concept of continuous type~\ref{en:casenolinkwrap} plaquette wrappings can be generalized from $\Un{1}$ to $\SU{N}$ plaquettes.
\changed{The relation between plaquette wrappings and the production of Dirac monopoles is discussed in Appendix~\ref{assec:situationinu1}.}

\subsection{Situation in $\SU{N}$}\label{ssec:situationinsun}

\changed{Compact Lie groups like $\SU{N}$ carry a Riemannian metric $g$ which is invariant under left and right translation. As a consequence, the well known bi-invariant Haar measure for group integration exists, which is given by the volume form of the invariant metric, normalized so that that the group volume is $1$. The fact that $\SU{N}$ has a Riemannian metric also means that we can at each point $U\in\SU{N}$ define a tangent space $\mathrm{T}_U\!\SU{N}$ and the so called exponential map $\exp_U\,:\,\mathrm{T}_U\!\SU{N}\to\SU{N}$ that maps vectors $\omega\in \mathrm{T}_U\!\SU{N}$ from the tangent space at point $U$ to points $V=\exp_{U}\of{\omega}$ in $\SU{N}$. The graph of the function $\gamma_{U,\omega}\of{t}=\exp_U\of{\omega\,t}$, $t\in\fof{0,1}$ then describes a geodesic segment in $\SU{N}$ that start in the point $U$ in the direction specified by $\omega$ and has geodesic length $\ssabs{\omega}_{g\of{U}}$.}

\changed{The exponential map $\exp_U\,:\,\mathrm{T}_U\!\SU{N}\to\SU{N}$ is bijective only in a neighborhood $\Omega_U\subset \mathrm{T}_U\!\SU{N}$ around the origin of $\mathrm{T}_U\!\SU{N}$. The boundary $\partial\Omega_U$ of the region $\Omega_U\subset\mathrm{T}_U\!\SU{N}$ over which the exponential map $\exp_U$ is bijective is called the cut locus of $U$ in $\mathrm{T}_U\!\SU{N}$ and its image $\exp_U\of{\partial\Omega_U}$ is, correspondingly, the cut locus of $U$ in $\SU{N}$.}

\changed{At the point $U=\id\in\SU{N}$ the tangent space to $\SU{N}$ corresponds to the usual Lie algebra $\su{N}=\mathrm{T}_{\id}\!\SU{N}$ and the exponential map applied to some $\omega\in\su{N}$ can be identified with the usual matrix exponential,
\[
\exp_{\id}\of{\omega}=\exp\of{\ii\,\omega^a\,T^{a}}\ ,
\]
with $\cof{T_{a}}_{a=1,\ldots,N^2-1}$ being a hermitian basis of $\su{N}$.}

\changed{The domain $\Omega_{\id}\subset \su{N}=\mathrm{T}_{\id}\!\SU{N}$ over which $\exp_{\id}$ is bijective is given by the set of hermitian matrices $\omega\in\su{N}$ whose eigenvalues $\cof{\phi_i}_{i=1,\ldots,N}$ satisfy~\cite{rossmann2002lie}:
\[
\abs{\phi_i}<\pi\quad\forall\,i\in\cof{1,\ldots,N}\ .\label{eq:sunbijectivitydomain}
\]
The image $\exp_{\id}\of{\partial\Omega_{\id}}\subset \SU{N}$, i.e. the cut locus of $\id$ in $\SU{N}$, is therefore given by the set of matrices $V\in\SU{N}$ for which at least one eigenvalue $\lambda_i=\e^{\ii\,\phi_i}$, $i=1,\ldots,N$ is equal to $-1$.}

\changed{To illustrate this, we pick a $\omega\in\su{N}$ and write it as
\[
\omega=U_{\omega} D_{\omega} U_{\omega}^{\dagger}\ ,
\]
with $U_{\omega}\in\SU{N}$ and $D_{\omega}=\operatorname{diag}\of{\phi_1,\ldots,\phi_N}$ being a diagonal matrix containing the eigenvalues of $\omega$. The image $V=\exp_{\id}\of{\omega}$ of $\omega$ under the exponential map $\exp_{\id}$ is then,
\[
V=U_{\omega} \exp\of{\ii\,D_{\omega}} U_{\omega}^{\dagger} = U_{\omega} D_{V} U_{\omega}^{\dagger}\ ,
\]
with $D_{V}=\operatorname{diag}\of{\lambda_1,\ldots,\lambda_N}$ being the diagonal matrix of eigenvalues of $V$, given by $\lambda_i=\e^{\ii\,\phi_i}$, $i\in\cof{1,\ldots,N}$. As long as $-\pi<\phi_i<\pi$ $\forall i\in\cof{1,\ldots,N}$, the exponential map is a bijection as the inverse map can be defined using,
\[
\phi_i=-\ii\,\log\of{\lambda_i}\quad,\quad i=1,\ldots,N\ ,\label{eq:pvlog}
\]
with $\log$ being the principal value complex logarithm. But, if $\abs{\phi_i}\geq\pi$ for any $i\in{1,\ldots,N}$, then \eqref{eq:pvlog} breaks down.}

\changed{We can now generalize the criterion we found in the previous section for the production of bulk configurations in $\Un{1}$ lattice gauge theory to the $\SU{N}$ case. In the $\Un{1}$ case we argued, that if the phase \eqref{eq:totplaqphaseu1} of any plaquette variable $U_{\mu\nu}\of{x}$ in the system crosses continuously the $\loint{-\pi,\pi}$-interval boundary (plaquette wrappping of type~\ref{en:nolinkwrap}), then a bulk configuration is produced. Such a continuous crossing of the $\loint{-\pi,\pi}$-interval boundary by the plaquette phase implies of course that the corresponding plaquette variable $U_{\mu\nu}\of{x}$ passes through the value $-1$, which is the cut locus for the exponential map from the Lie algebra $\mathrm{u}\of{1}=\mathrm{T}_{1}\!\Un{1}=\mathbb{R}$ to the group manifold $\Un{1}$. If a plaquette variable $U_{\mu\nu}\of{x}$ approaches and smoothly crosses the cut locus value $-1$ the resulting lattice gauge field described by the link variables can no longer be properly mapped on a corresponding Lie algebra valued gauge field that is relevant for continuum physics. The same applies in the case of $\SU{N}$ lattice gauge theory if a $\SU{N}$ plaquette variable approaches and smoothly cross the $\SU{N}$ cut locus. The resulting configuration cannot be mapped on a corresponding Lie algebra valued gauge configuration that is relevant for continuum physics, which qualifies it as a bulk configuration.}

\changed{In order to prevent a $\SU{N}$ gauge system form entering a bulk phase at strong coupling, the plaquette variables must not continuously cross the cut locus. As discussed below \eqref{eq:sunbijectivitydomain}, this means that the eigenvalues of the plaquette variables must not continuously cross the value $-1$.}

\changed{The question of how such a restriction affects the creation of magnetic monopoles in $\SU{N}$ gauge theory is addressed in Appendix~\ref{assec:situationinsun}.}

\subsection{Bulk-preventing action}\label{ssec:bulkprevaction}
To prevent plaquettes from having eigenvalues close to $-1$, we introduce the following family of gauge actions:
\[
S_{G,b}\,=\,\frac{2\,\gamma}{n\,N}\sum\limits_{x}\sum\limits_{\mu<\nu}\trace\sof{\sof{\Omega^{\dagger}_{\mu\nu}\of{x}\Omega_{\mu\nu}\of{x}}^{-n}-\id}\ ,\label{eq:bpaction}
\]
with $n\geq 1$ and
\[
\Omega_{\mu\nu}\of{x}=\sof{\id+\underbrace{U_{\mu\nu}\of{x}}_{\mathclap{\text{plaquette}}}}/2 .\label{eq:omegamat}
\]
The form of the actions \eqref{eq:bpaction} was inspired by the dislocation-prevention action introduced in~\cite{DeGrand:2014rwa}. The naive continuum limit of \eqref{eq:bpaction} is the same as for the Wilson gauge action $S_{G,W}$. This can be seen by writing $U_{\mu\nu}\of{x}=\exp\sof{\ii\,{s}\,F'_{\mu\nu}\of{x}}$, with ${s}\,F'_{\mu\nu}={a^2}\,F_{\mu\nu}+\order{a^3}$ and expanding in a power series in $s$. For the local plaquette action contributing to $S_{G,W}$ from \eqref{eq:wilsongaugeaction} one then finds:
\begin{multline}
\repart\trace\of{\id-U_{\mu\nu}\of{x}}={s^2} \trace\of{F'_{\mu\nu}\of{x}F'_{\mu\nu}\of{x}}/2\\
-{s^4}\changedn{\frac{1}{12}\ucases{\sof{\trace\sof{F'_{\mu\nu}\of{x}F'_{\mu\nu}\of{x}}/2}^2 &\text{if}\ N<4\\\ \,\trace\sof{\sof{F'_{\mu\nu}\of{x}F'_{\mu\nu}\of{x}}^2}/2 &\text{if}\ N\geq 4}}\\
+\order{s^6}\ ,\label{eq:expansionwilsonaction}
\end{multline}
and, correspondingly, for the local \changed{plaquette} action contributing to $S_{G,b}$ from \eqref{eq:bpaction}:
\begin{multline}
\frac{2}{n}\trace\sof{\sof{\Omega^{\dagger}_{\mu\nu}\of{x}\Omega_{\mu\nu}\of{x}}^{-n}-\id}={s^2}\trace\of{F'_{\mu\nu}\of{x}F'_{\mu\nu}\of{x}}/2\\
+{s^4}\changedn{\frac{1+3\,n}{24}\ucases{\sof{\trace\sof{F'_{\mu\nu}\of{x}F'_{\mu\nu}\of{x}}/2}^2 &\text{if}\ N<4\\\ \,\trace\sof{\sof{F'_{\mu\nu}\of{x}F'_{\mu\nu}\of{x}}^2}/2 &\text{if}\ N\geq 4}}\\
+\order{{s^6}}\ .\label{eq:expansionbpaction}
\end{multline}
In the limit $\of{a\to 0}$, one has $s\sim a^2$, $F'_{\mu\nu}\of{x}\sim F_{\mu\nu}\of{x}$, and we see that the two local actions have the same leading term~$\sim a^4$, namely:
\[
a^4\,\trace\of{F_{\mu\nu}\of{x}F_{\mu\nu}\of{x}}/2\ .
\]

The actions \eqref{eq:bpaction} introduce an infinite potential barrier between bulk and continuum configurations. This is sufficient to ensure that, if we start a simulation from a cold configuration (all link variables equal to the identity) and use a hybrid Monte Carlo (HMC) algorithm to update the gauge system, no bulk-configurations will be produced.

One could infer that our bulk-preventing setup results in a non-ergodic update algorithm. However, one should keep in mind that the part of the configuration space that is not sampled is irrelevant for the continuum limit of the theory. The algorithm prevents ensemble averages of the lattice system from being contaminated (or even dominated) by bulk-configurations, which should allow one to extract continuum physics also at stronger coupling. The same effect could be achieved by defining a modified measure, which gives zero weight to bulk configurations.  However, this would be difficult to implement as the latter are hard to identify once they are created. The use of an action \eqref{eq:bpaction} in combination with an HMC algorithm is a proxy to achieve the same effect but in a simpler and more economic way. An expression for the gauge force, required for the HMC, is presented in Appendix~\ref{asec:compgaugeforce}. 

Let us note that for $\Un{1}$ and $\SU{2}$, the actions in \eqref{eq:bpaction} alone would have a similar effect as the topological actions discussed in \cite{Brower:1982ak,Bornyakov:1991gq,Akerlund:2015zha,Nogradi:2018ivi}: the larger the inverse bare coupling $\gamma$, the stronger the plaquette values are repelled from $-1$ resp. $-\id$. For $\SU{N}$ with $N>2$ the effect of \eqref{eq:bpaction} is, however, different from the one of the topological action discussed in \cite{Nogradi:2018ivi}, as for $N>2$ the trace of the plaquette does no-longer completely determine the plaquette eigenvalues. An action which can have a similar effect as our actions \eqref{eq:bpaction} has been given in \cite{Brandt:2019ukx}. \changed{We also note that based on the observation that the bulk transition of the Wilson gauge action \eqref{eq:wilsongaugeaction} is  sensitive to the addition of an adjoint plaquette term~\cite{Bhanot:1981eb}, it has in~\cite{Hasenfratz:2011xn} been demonstrated that for $N=3$ the bulk phase can be avoided by adding an appropriately scaled negative adjoint plaquette term to the standard Wilson gauge action~\cite{Hasenbusch:2004yq,Hasenfratz:2011xn}}. 

While for us it is desirable that the actions \eqref{eq:bpaction} do not prevent topology from fluctuating, there have also been attempts to find actions which keep the topology fixed \cite{Bietenholz:2005rd}. 
\FloatBarrier

\section{Results}\label{sec:results}

\begin{figure*}[!htb]
\begin{minipage}[t]{0.3\linewidth}
\vspace{0pt}
\centering
\includegraphics[height=0.96\linewidth,keepaspectratio,right]{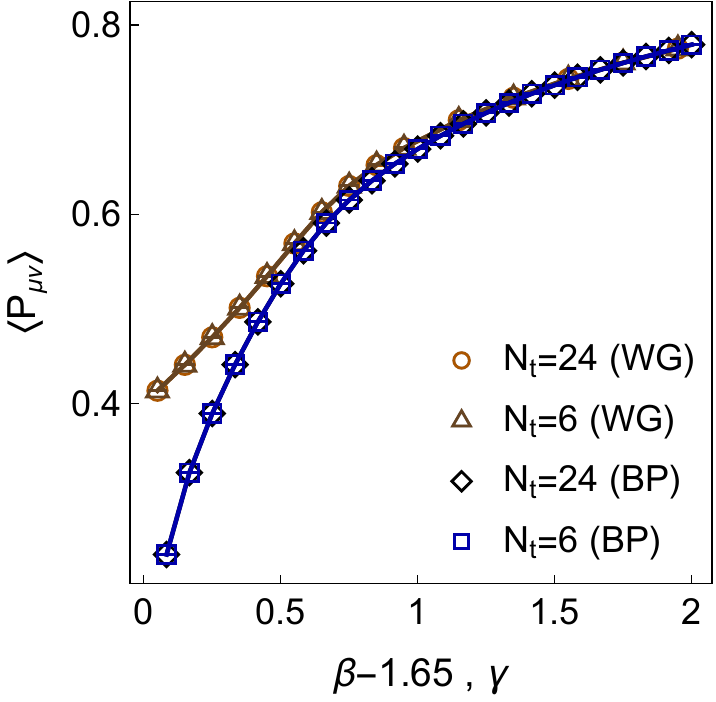}
\end{minipage}\hfill
\begin{minipage}[t]{0.3\linewidth}
\vspace{0pt}
\centering
\includegraphics[height=0.96\linewidth,keepaspectratio,right]{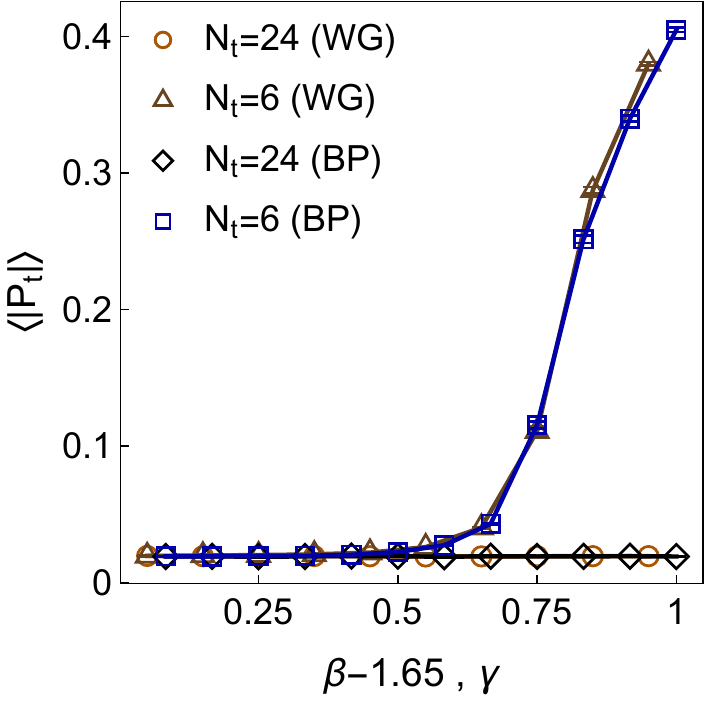}
\end{minipage}\hfill
\begin{minipage}[t]{0.3\linewidth}
\vspace{0pt}
\centering
\includegraphics[height=0.96\linewidth,keepaspectratio,right]{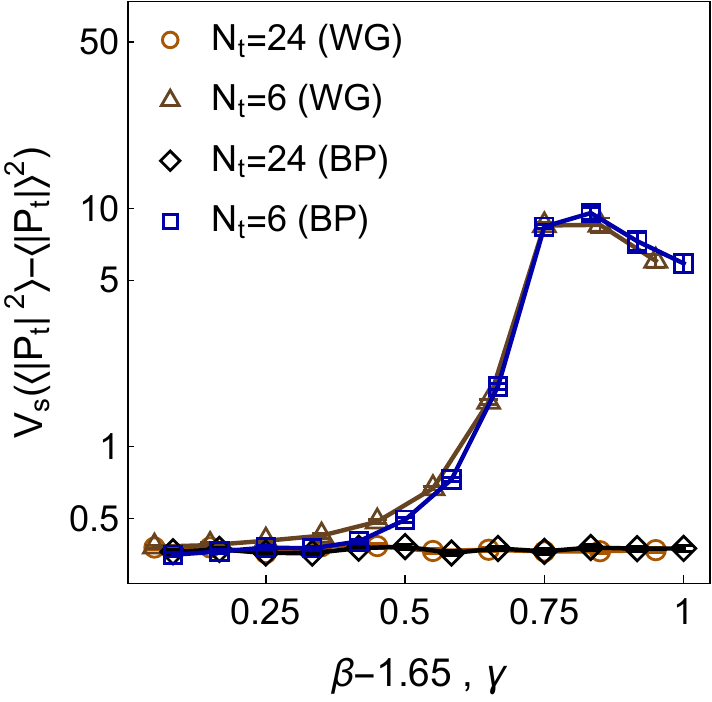}
\end{minipage}\\[7pt]
\begin{minipage}[t]{0.3\linewidth}
\vspace{0pt}
\centering
\includegraphics[height=0.96\linewidth,keepaspectratio,right]{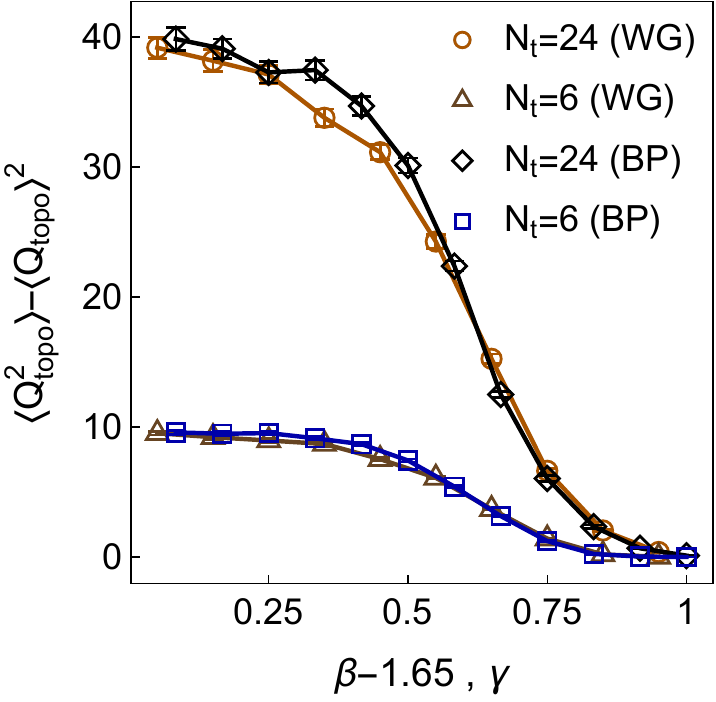}
\end{minipage}\hfill
\begin{minipage}[t]{0.3\linewidth}
\vspace{0pt}
\centering
\includegraphics[height=0.96\linewidth,keepaspectratio,right]{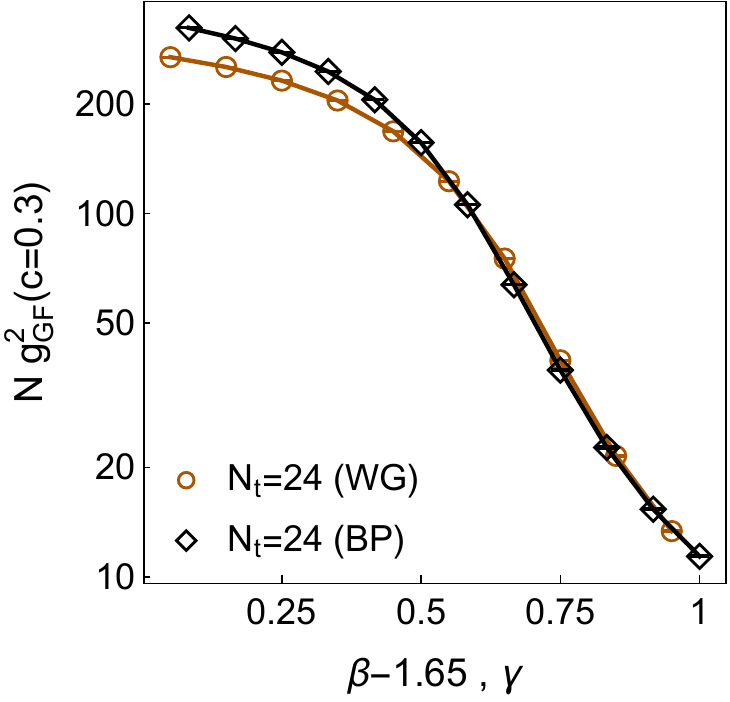}
\end{minipage}\hfill
\begin{minipage}[t]{0.3\linewidth}
\vspace{0pt}
\centering
\includegraphics[height=0.96\linewidth,keepaspectratio,right]{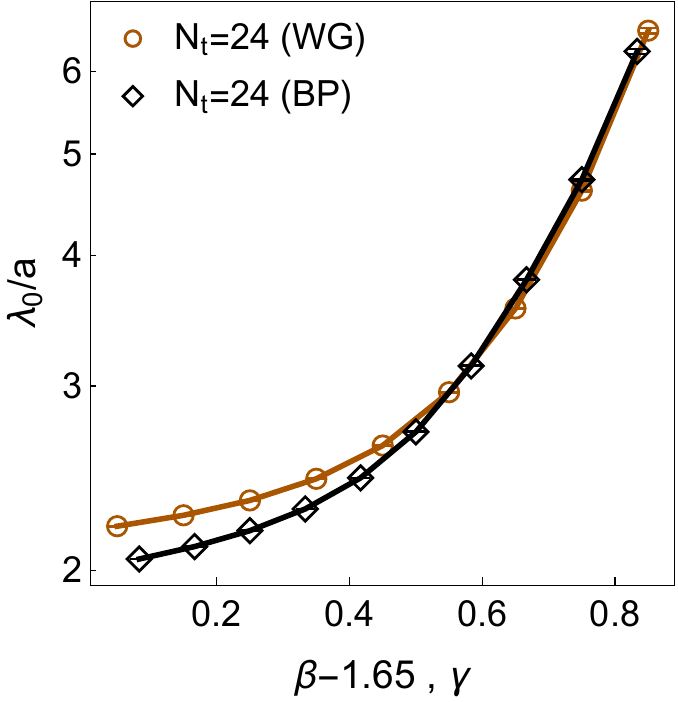}
\end{minipage}
\caption{Comparison of pure $\SU{2}$ gauge theory results obtained with the Wilson gauge action (WG), Eq.~\eqref{eq:wilsongaugeaction} (orange circles and brown triangles) and the bulk-preventing (BP) action, Eq.~\eqref{eq:bpaction}, for $n=2$ (black diamonds and blue squares). To guide the eye, the data points are connected by straight lines. The first row shows the real part of the average traced plaquette (top left), the temporal Polyakov loop (top center), and the temporal Polyakov loop variance (top right). The second row shows the topological susceptibility (bottom left) the gradient flow coupling at $c=\lambda/N_s=0.3$ (bottom center), with flow scale $\lambda=\sqrt{8\,t}$, corresponding to flow time $t$, and $\lambda_0/a$ (bottom right), which is the flow scale $\lambda$ (in lattice units) at which \changed{$t^2\,\avof{E\of{t}}=0.26$} with $\avof{E\of{t}}$ being the clover action density of the flowed gauge field at flow time $t$. On our finite lattices with spatial size $V_s=12^3$, the temporal size is set to $N_t=6$ for finite temperature, and to $N_t=24$ for zero-temperature. The results are shown as functions of $\beta-1.65$ (WG) resp. $\gamma$ (BP).}
\label{fig:actioncompsu2}
\end{figure*}

To test whether the bulk-preventing actions~\eqref{eq:bpaction} deserve their name and whether they are able to reproduce the same weak coupling results as the Wilson gauge action, we carried out simulations with pure gauge $\SU{2}$, pure gauge $\SU{5}$, and $\SU{3}$ with $N_{f}=4$ Wilson fermion flavours. With the Wilson gauge action, all three of these theories enter a bulk phase for sufficiently small values of the inverse gauge coupling $\beta$. For pure gauge $\SU{2}$, the transition is a smooth cross-over, while for pure gauge $\SU{5}$ and for the fermionic $\SU{3}$ theory with sufficiently large fermion hopping parameter $\kappa$ the transition is of first order. In the following we will discuss the three cases separately. We use the version of the action \eqref{eq:bpaction} with $n=2$. The choice of $n$ should not affect physical results, but it turns out that a too small value of $n$ will require also a smaller step size in the HMC trajectories to achieve similar acceptance rates, which can become computationally more expensive than using $n=2$.

According to the expansions \eqref{eq:expansionwilsonaction} and \eqref{eq:expansionbpaction}, the Wilson gauge (WG) action \eqref{eq:wilsongaugeaction} and bulk-preventing (BP) action \eqref{eq:bpaction} agree only to order \changed{$s^2\sim a^4$}. Thus, the inverse bare couplings $\beta$ and $\gamma$ will not be equal in the weak coupling limit. However, it turns out that locally the two couplings can be related quite accurately by a constant shift, $\gamma=\beta-\Delta\beta$, which we will use to directly compare bare lattice results obtained with the two different actions. Of course, there is in general no need for bare lattice results, obtained with different actions, to agree. However, it seems that in the present case, the systems controlled by the WG and the BP action  behave in the weak coupling regime sufficiently equally, so that a direct comparison of bare lattice results is reasonable. 

\subsection{$\SU{2}$ pure gauge}\label{ssec:puregaugesu2results}

$\SU{2}$ pure gauge theory with the WG action \eqref{eq:wilsongaugeaction} is known to have a smooth cross-over between the bulk phase and the continuum-like phase. Thus, for this theory the BP action \eqref{eq:bpaction} is not expected to provide any significant advantage over the WG action and both actions should give rise to the same results not just at weak, but also all the way down to strong coupling.
 
In Fig.~\ref{fig:actioncompsu2} we compare results obtained with the WG and the BP action. The WG data is plotted as function of $\beta-\Delta\beta$ with $\Delta\beta=1.65$ and the BP data is plotted as function of $\gamma$. The shift $\Delta\beta=1.65$ has been determined by requiring that the "spatial deconfinement" transition, at which the spatial Polyakov loop develops a non-zero expectation value (indicating that the physical spatial volume becomes too small to fit a meson), occurs for the two actions at the same value of $(\beta-\Delta\beta)$ resp.~$\gamma$. 

The top-left panel in Fig.~\ref{fig:actioncompsu2} shows the average of the traced plaquette and we note that when plotted against $\beta-1.65$ resp. $\gamma$, the plaquette values for the two different actions agree remarkable well at sufficiently weak coupling. Only below $\gamma=\beta-1.65\approx 0.65$, where the WG action enters the bulk-phase, the plaquette value for the WG action starts to deviate from the BP one as function of $\gamma=\beta-1.65$. 

The strong coupling limit is for both actions obtained by sending their respective inverse coupling to zero, i.e. $\of{\gamma\to 0}$ and $\of{\beta\to 0}$. With the WG action the strong coupling phase extends over the interval $\beta\in\fof{0,2.3}$, while with the BP action, the strong coupling phase extends over the significantly smaller interval $\gamma\in\fof{0,0.65}$. In the strong coupling phase, the system should therefore with the WG action change more slowly as function of $\beta$ than with the BP action as function of $\gamma$. 

This is indeed what can be observed in the remaining panels of Fig.~\ref{fig:actioncompsu2}: the results obtained with the two actions for temporal Polyakov loop (top-center), Polyakov loop variance (top-right), and topological susceptibility (bottom-left) are consistent and match for $\gamma=\beta-1.65>0.65$ very nicely as functions of $\gamma$ resp. $\beta-1.65$, while for $\gamma=\beta-1.65<0.65$, the WG results change more slowly as function of $\beta$ than the corresponding BP results do as function of $\gamma$.

The last two panels on the second row of Fig.~\ref{fig:actioncompsu2} show gradient flow quantities: the bottom-center panel shows the gradient flow coupling, $N\,g_{\mathrm{GF}}^2\of{c}$ at $c=\lambda/N_s=0.3$, with $\lambda=\sqrt{8\,t}$ being the flow scale corresponding to flow time $t$; and the bottom-right panel shows $\lambda_0/a$, which is the flow scale $\lambda$ (in lattice units) at which \changed{$t^2\,\avof{E\of{t}}=0.26$~\cite{DeGrand:2017gbi}} with $E\of{t}$ being the clover action density of the flowed gauge field at flow time $t$. Both gradient flow quantities have been corrected for leading finite volume and lattice spacing effects~\cite{Rindlisbacher:2021hhh} by replacing in the computation of the continuum finite volume correction given in~\cite{Fodor:2012td} the continuum momenta by lattice momenta and adapting the formula to the case $N_t\neq N_s$. \changed{A refined correction formula that takes into account the details of the utilized lattice actions for sampling and flowing of the gauge fields, as well as for measuring $E\of{t}$, is presented in~\cite{Fodor:2014cpa}}. Also for the gradient flow quantities the data obtained with the two actions agrees as function of $\gamma=\beta-1.65$ if $\gamma=\beta-1.65>0.65$, whereas for $\gamma=\beta-1.65<0.65$, the WG data changes more slowly as function of $\beta$ than the BP data does as function of $\gamma$.

As mentioned at the beginning of this section, the shift $\Delta\beta=1.65$ has been determined by requiring that the spatial deconfinement happens at the same value of $\gamma$ and $\beta-\Delta\beta$. Because of our small spatial volumes of linear size $N_s=12$, this happens already at $\gamma=\beta-1.65 > 0.8$.

\FloatBarrier

\subsection{$\SU{5}$ pure gauge}\label{ssec:puregaugesu5results}

\begin{figure*}[htb]
\begin{minipage}[t]{0.3\linewidth}
\vspace{0pt}
\centering
\includegraphics[height=0.96\linewidth,keepaspectratio,right]{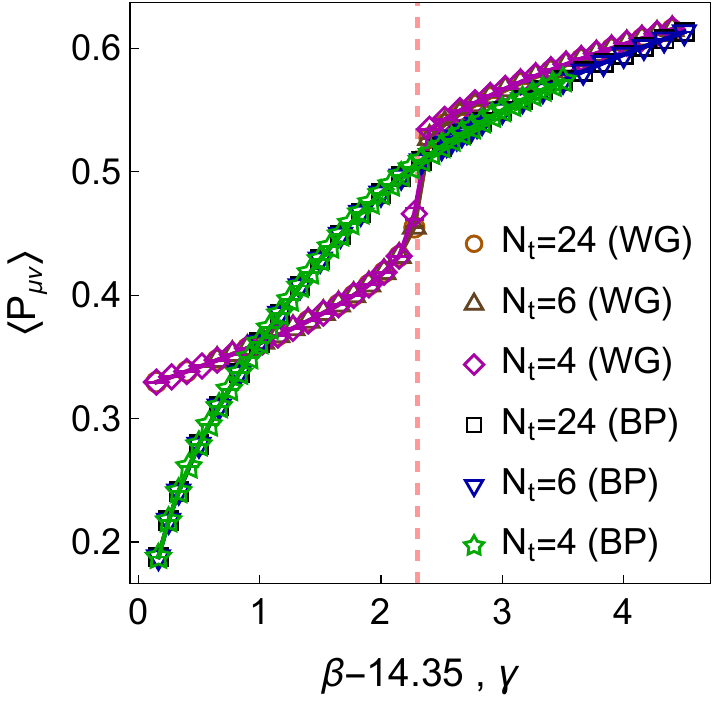}
\end{minipage}\hfill
\begin{minipage}[t]{0.3\linewidth}
\vspace{0pt}
\centering
\includegraphics[height=0.96\linewidth,keepaspectratio,right]{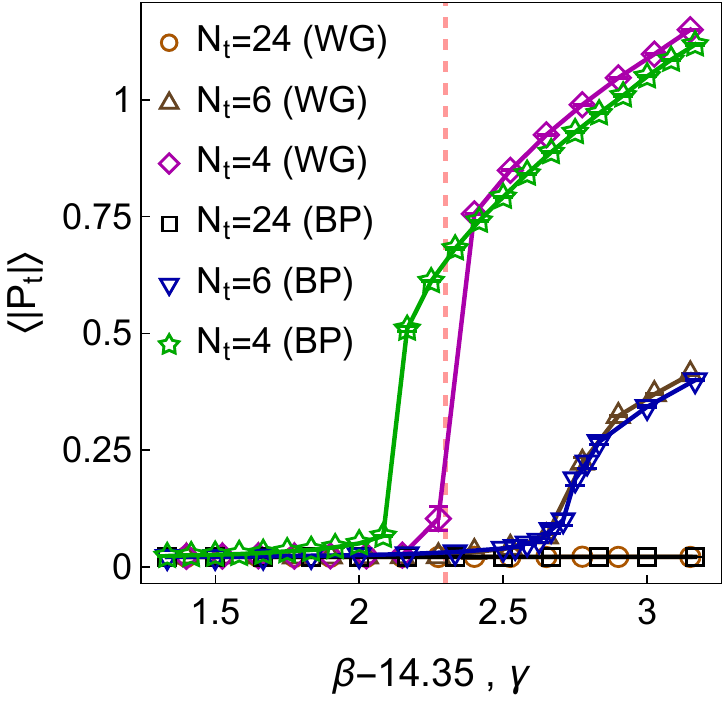}
\end{minipage}\hfill
\begin{minipage}[t]{0.3\linewidth}
\vspace{0pt}
\centering
\includegraphics[height=0.96\linewidth,keepaspectratio,right]{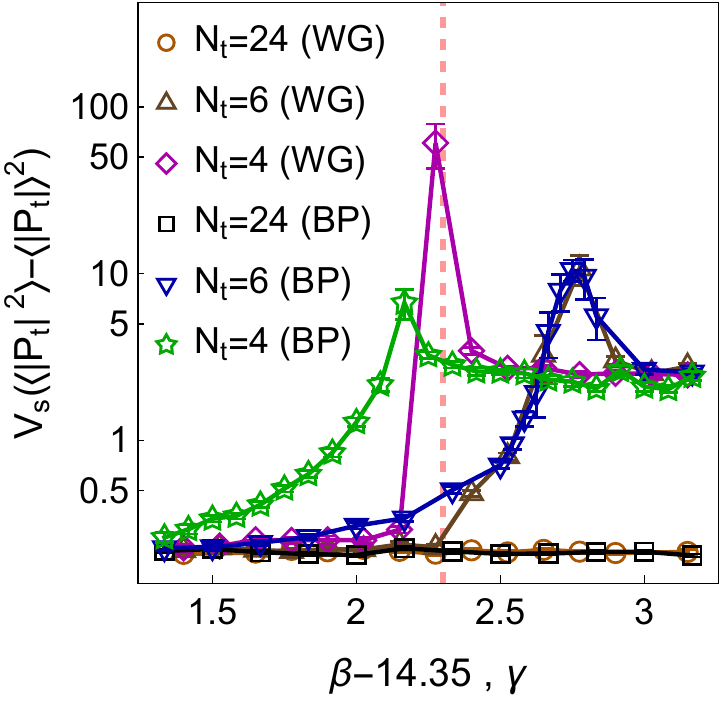}
\end{minipage}\\[7pt]
\begin{minipage}[t]{0.3\linewidth}
\vspace{0pt}
\centering
\includegraphics[height=0.96\linewidth,keepaspectratio,right]{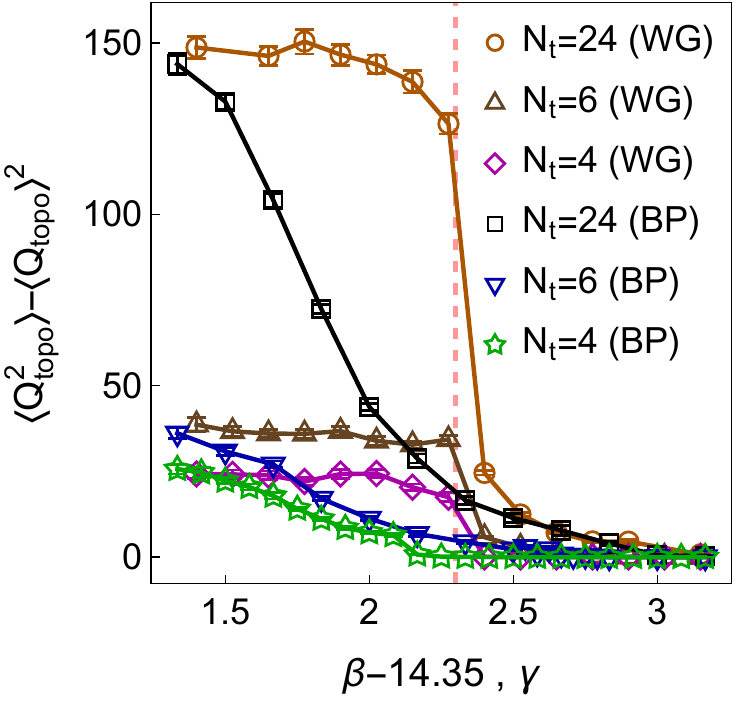}
\end{minipage}\hfill
\begin{minipage}[t]{0.3\linewidth}
\vspace{0pt}
\centering
\includegraphics[height=0.96\linewidth,keepaspectratio,right]{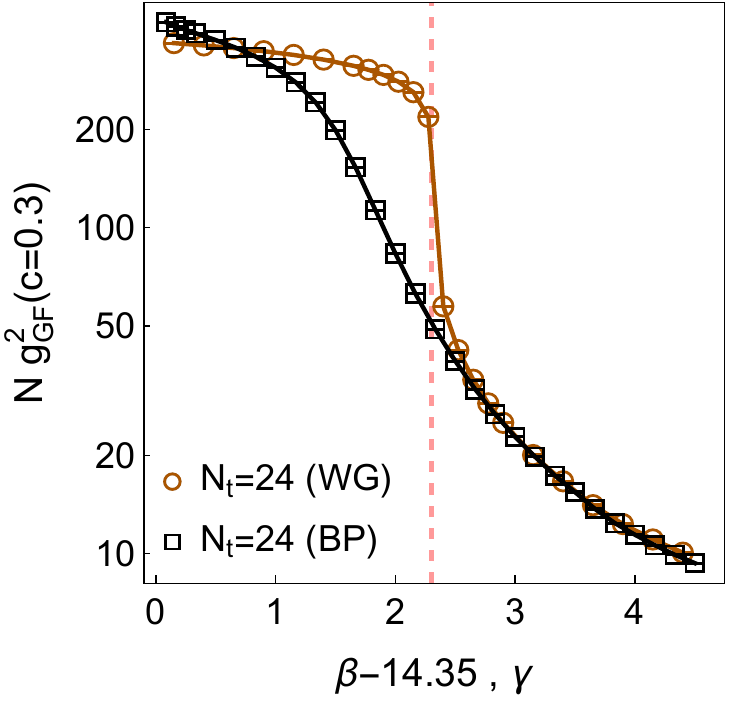}
\end{minipage}\hfill
\begin{minipage}[t]{0.3\linewidth}
\vspace{0pt}
\centering
\includegraphics[height=0.96\linewidth,keepaspectratio,right]{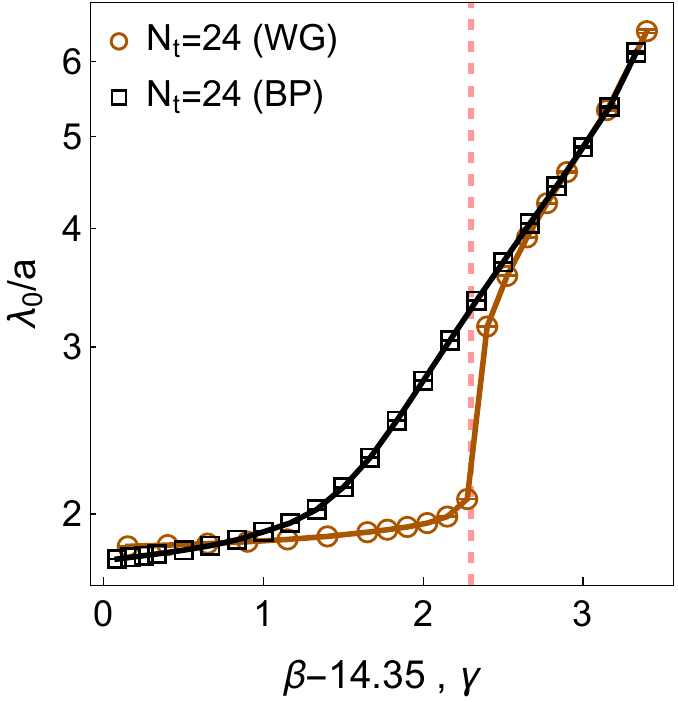}
\end{minipage}
\caption{Same as Fig.~\ref{fig:actioncompsu2}, but for pure gauge $\SU{5}$ and including the case $N_t=4$ for which the WG action can no longer properly resolve the finite temperature transition as the latter is forced to occur on top of the bulk transition. The dashed vertical red line indicates the approximate location of the bulk transition of the WG action. Note that the shown data was obtained on small lattices with $N_s=12$ and we did not attempt to perform simulations directly at the pseudo critical points. The peaks visible in the data for the Polyakov loop variance (top-right panel) do therefore not reflect the true pseudocritcal behavior; the lines simply connect the available data points to guide the eye.}
\label{fig:actioncompsu5}
\end{figure*}

Fig.~\ref{fig:actioncompsu5} provides the same information as Fig.~\ref{fig:actioncompsu2} but for $\SU{5}$ instead of $\SU{2}$. The shift in $\beta$ required to match $\gamma$ at weak coupling has been set to $\Delta\beta=14.35$, which, as in the $\SU{2}$-case, is determined by matching the values of $\gamma$ and $\beta-\Delta\beta$ at which spatial deconfinement occurs.

For $\SU{5}$ the bulk transition of the WG action is of 1st order~\cite{Lucini:2005vg}, which is clearly visible from the sharp discontinuity in the WG data for the average plaquette at $\beta-14.35\approx 2.3$, shown in the top-left panel of Fig.~\ref{fig:actioncompsu5}. 
In contrast, with the BP action the plaquette is completely continuous as a function of the inverse gauge coupling $\gamma$. In the continuum phase, i.e. for $\gamma=\beta-14.35>2.3$, the average plaquette values obtained with the two different actions converge only slowly with increasing inverse coupling, while the average temporal Polyakov loop (top-center), temporal Polyakov loop variance (top-right) and topological susceptibility (bottom-left), as well as the gradient flow quantities, $N\,g_{\mathrm{GF}}^{2}\of{c=0.3}$ (bottom-center) and $\lambda_0/a$ (bottom-right), agree for the two actions almost immediately when $\gamma=\beta-14.35>2.3$.

The finite temperature transition for $N_t=6$ occurs at around $\gamma=\beta-14.35\approx 2.9$. This is above the value at which the WG action undergoes the bulk transition, and the properties of the finite temperature transition should therefore be described equally well with the WG and with the BP action. The temporal Polyakov loop (top-center) looks indeed the same for the two actions at $N_t=6$, and also the temporal Polyakov loop variance (top-right) agrees very well for $\gamma=\beta-14.35>2.3$ at $N_t=6$; at $\gamma=\beta-14.35\approx 2.3$ (red vertical dashed line) one can, however, notice a small jump in the Polyakov loop variance for the WG action, while the Polyakov loop variance obtained with the BP action behaves completely regular across this point.

A similar behavior can be observed in the $N_t=6$ data for the topological susceptibility (bottom-left), where the results obtained with the WG and BP action agree for $\gamma=\beta-14.35> 2.3$, but as $\gamma=\beta-14.35$ decreases below the bulk transition point, $\gamma=\beta-14.35\approx 2.3$, the topological susceptibility obtained with the WG action jumps and approaches almost immediately its strong-coupling plateau value, while with the BP action, the topological susceptibility approaches its strong coupling value much more smoothly.

With $N_t=4$, the WG action is no longer able to properly resolve the finite temperature transition. The data obtained with the BP action suggests that the finite temperature transition should for $N_t=4$ occur at $\gamma=\beta-14.35\approx 2.1$. With the WG action, the system is in the bulk-phase at this value of the bare gauge coupling~\cite{Lucini:2005vg}. It appears that the finite temperature transition cannot take place inside the bulk phase and occurs therefore on top of the bulk transition. Also the topological susceptibility obtained with the WG action for $N_t=4$ appears to be unable to decrease as long as the system is in the bulk phase. As in the case of $N_t=6$, the topological susceptibility obtained with the WG action appears also for $N_t=4$ to be stuck at the strong coupling plateau value for $\gamma=\beta-14.35<2.3$ and to decrease abruptly at $\gamma=\beta-14.35\approx 2.3$ when the inverse coupling is increased beyond this point. In contrast, with the BP action the asymptotic strong coupling value of the topological susceptibilities is also for $N_t=4$ approached smoothly.

The measurements of the gradient flow coupling $N\,g_{\mathrm{GF}}^2\of{c}$ at $c=\lambda/N_s=0.3$ are shown in the bottom-center panel of Fig.\ref{fig:actioncompsu5}. From this we conclude that the WG action is not capable of reaching gradient flow couplings larger than $g_{\mathrm{GF}}^2 \approx 11$ before hitting the bulk transition. In the bottom-right panel we show the flow scale $\lambda_0/a$ at which \changed{$t^2\,\avof{E\of{t}}=0.47$~\cite{DeGrand:2017gbi}}. For the WG action there is a discontinuity in $\lambda_0/a$ at the bulk transition point, indicating that there is a largest reachable lattice spacing. For the BP action these problems disappear and the gradient flow quantities behave smoothly. As in the $\SU{2}$ case, both gradient flow quantities have been corrected for finite volume and finite lattice spacing effects.
\FloatBarrier

\subsection{$\SU{3}$ with $N_{f}=4$ Wilson fermions}\label{ssec:fermionicresults}

For a $\SU{3}$ lattice gauge theory with the Wilson gauge action, the transition between continuum- and bulk-phase is normally a cross-over. However, if the theory is coupled to fermions, the transition can turn 1st order. 

\begin{figure}[htb]
\centering
\includegraphics[width=0.8\linewidth]{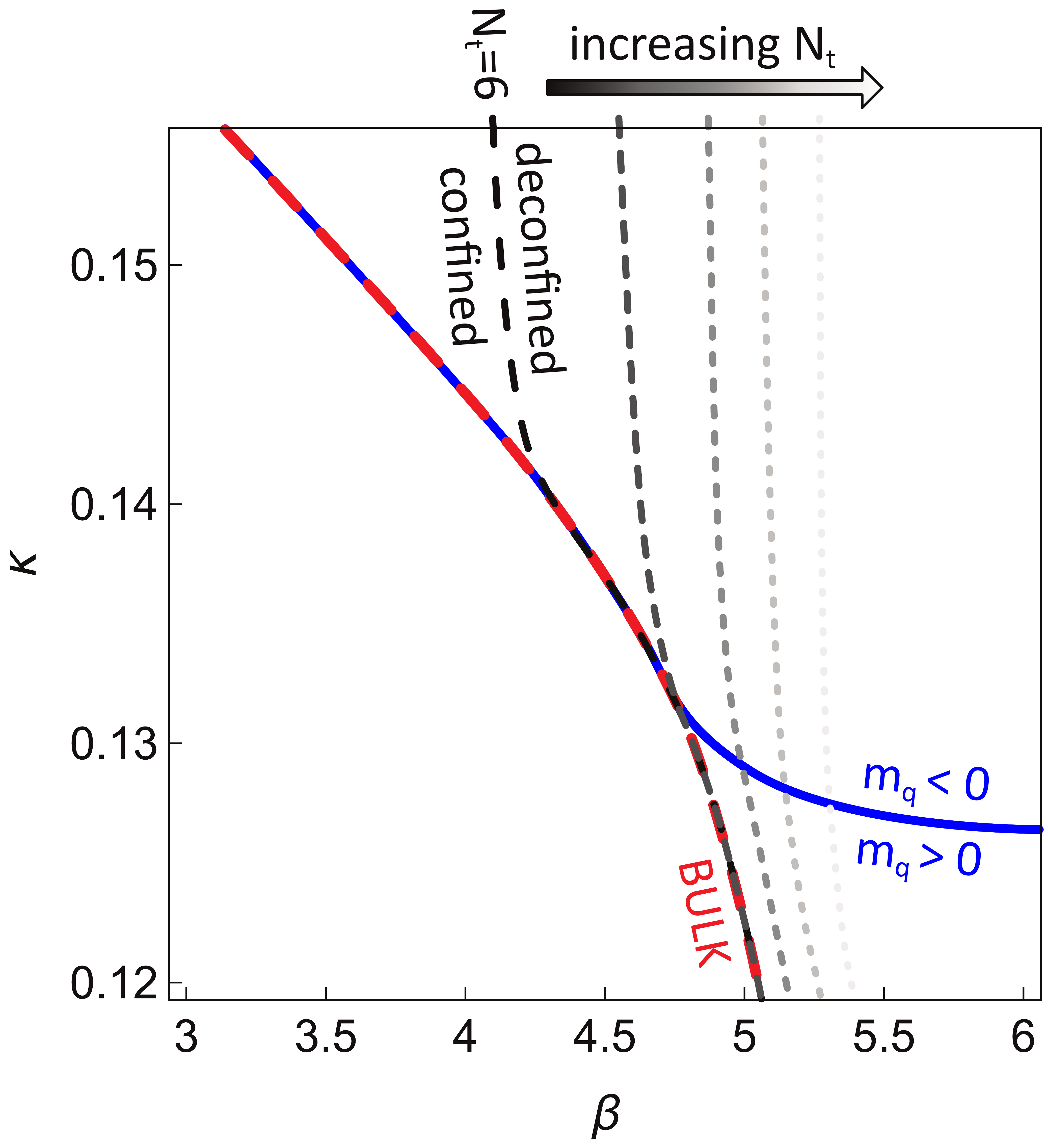}
\caption{Sketch of the finite temperature phase diagram for $\SU{3}$ lattice gauge theory with Wilson gauge action and $N_f=4$ Wilson clover fermion flavors (coupling to the gauge field via two-step stout smeared gauge links). The parameters $\beta$ and $\kappa$ are, respectively, the inverse gauge coupling and fermion hopping parameter. The red, dashed line marks the location of bulk transition (resp. crossover if $\kappa$ is sufficiently small) of the Wilson gauge action, the blue line shows where the PCAC quark mass, $m_q$, vanishes, and the dashed lines in different shades of gray mark the location of the finite temperature "deconfinement" transition lines for different values of $N_t$. The locations of the bulk transition and the $m_q=0$ line were estimated with simulations on a $12^3\times 24$ lattice, and the $N_t=6$ "deconfinement" transition line from simulations on a $12^3\times 6$ lattice. Note that the $N_f=4$ fermions cause the pseudo-critical $\beta$ to be lower than it would be in the pure-gauge case.}
\label{fig:phasediagramsu3}
\end{figure}

\begin{figure*}[htb]
\begin{minipage}[t]{0.3\linewidth}
\vspace{0pt}
\centering
\includegraphics[height=0.96\linewidth,keepaspectratio,right]{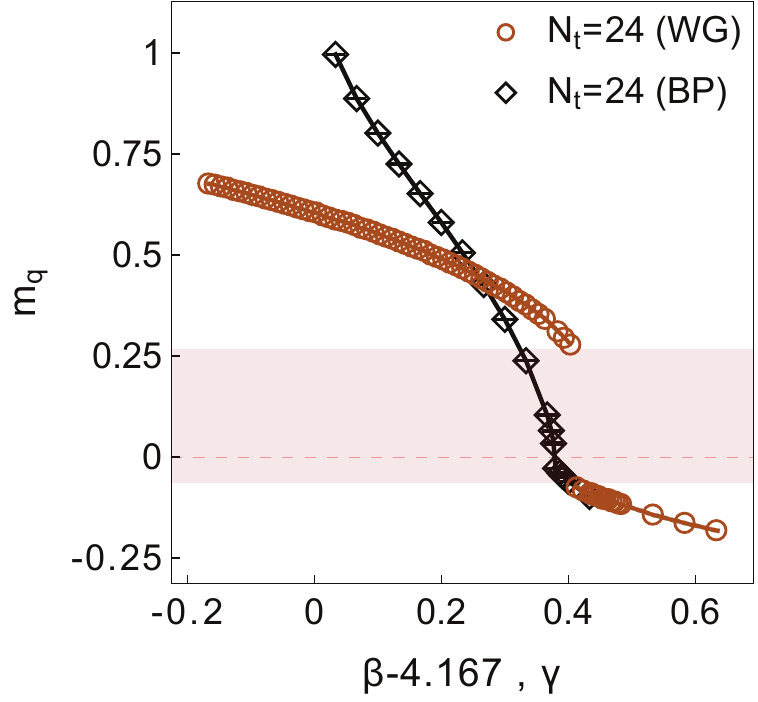}
\end{minipage}\hfill
\begin{minipage}[t]{0.3\linewidth}
\vspace{0pt}
\centering
\includegraphics[height=0.96\linewidth,keepaspectratio,right]{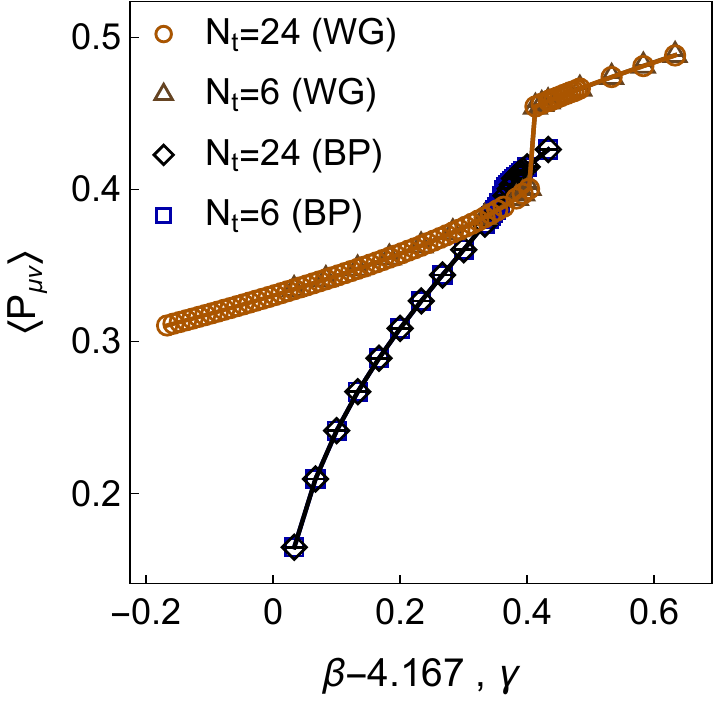}
\end{minipage}\hfill
\begin{minipage}[t]{0.3\linewidth}
\vspace{0pt}
\centering
\includegraphics[height=0.96\linewidth,keepaspectratio,right]{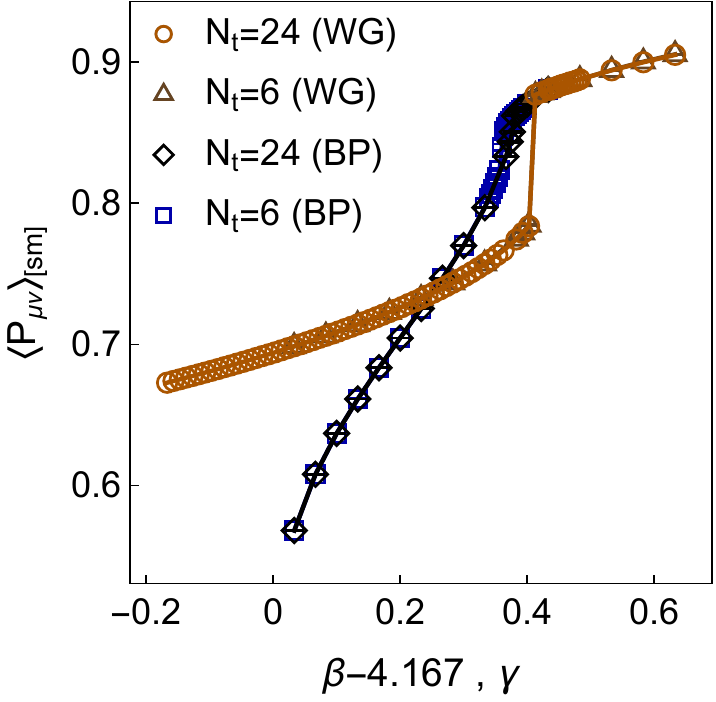}
\end{minipage}\\[7pt]
\begin{minipage}[t]{0.3\linewidth}
\vspace{0pt}
\centering
\includegraphics[height=0.96\linewidth,keepaspectratio,right]{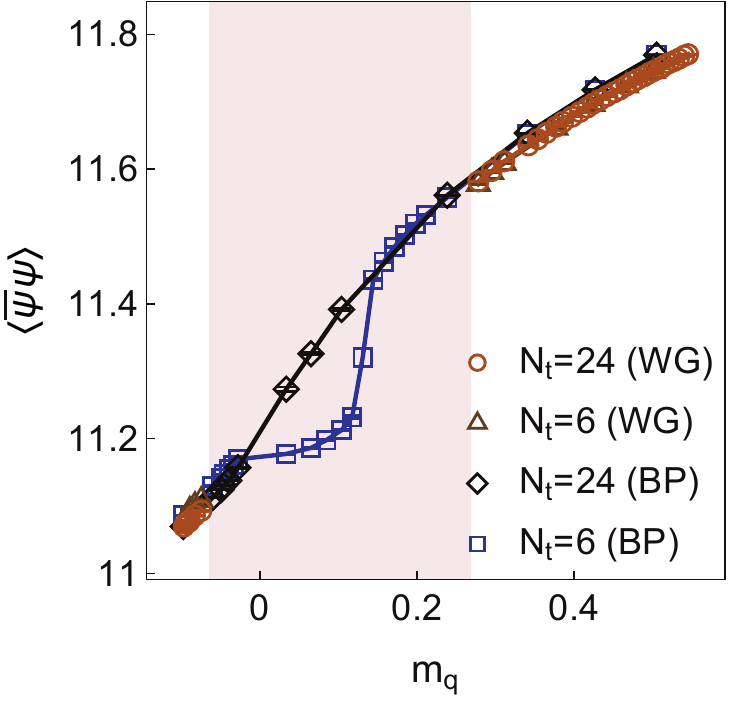}
\end{minipage}\hfill
\begin{minipage}[t]{0.3\linewidth}
\vspace{0pt}
\centering
\includegraphics[height=0.96\linewidth,keepaspectratio,right]{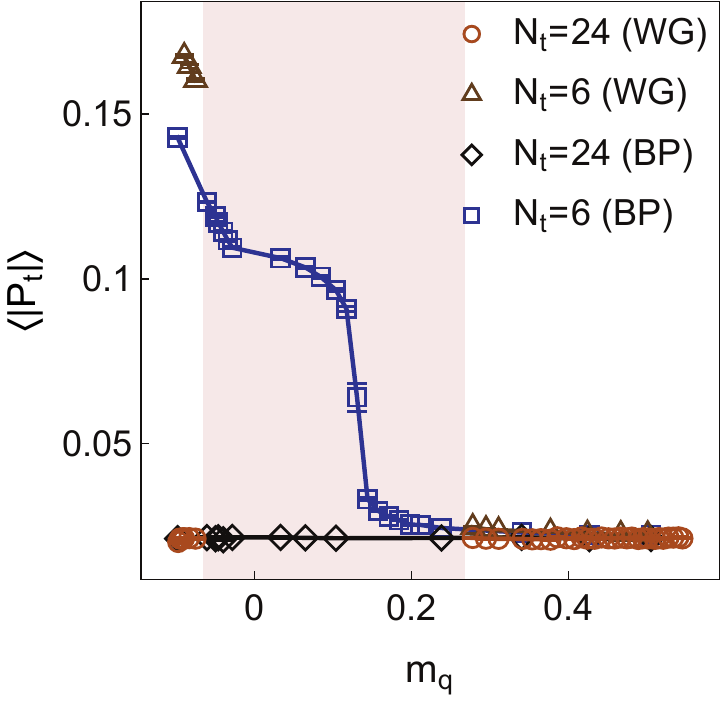}
\end{minipage}\hfill
\begin{minipage}[t]{0.3\linewidth}
\vspace{0pt}
\centering
\includegraphics[height=0.96\linewidth,keepaspectratio,right]{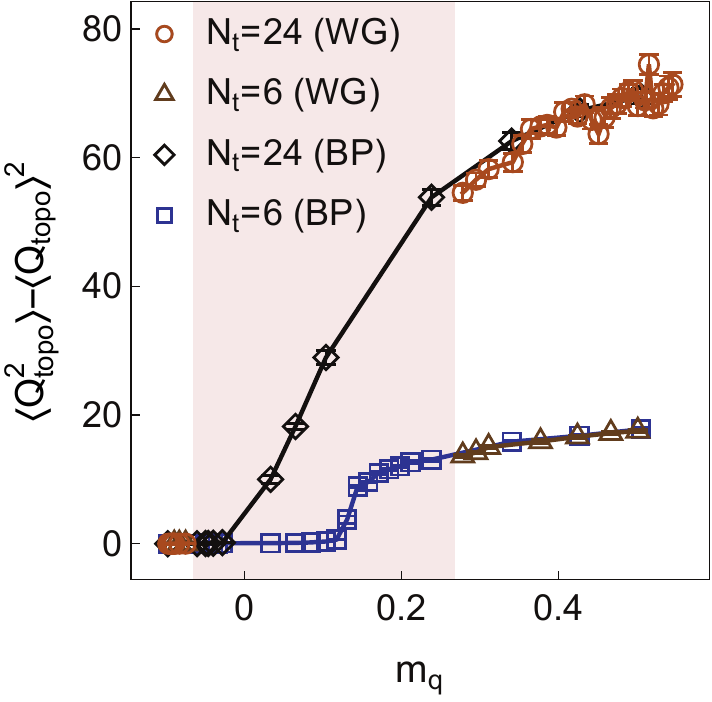}
\end{minipage}\\[7pt]
\begin{minipage}[t]{0.3\linewidth}
\vspace{0pt}
\centering
\includegraphics[height=0.96\linewidth,keepaspectratio,right]{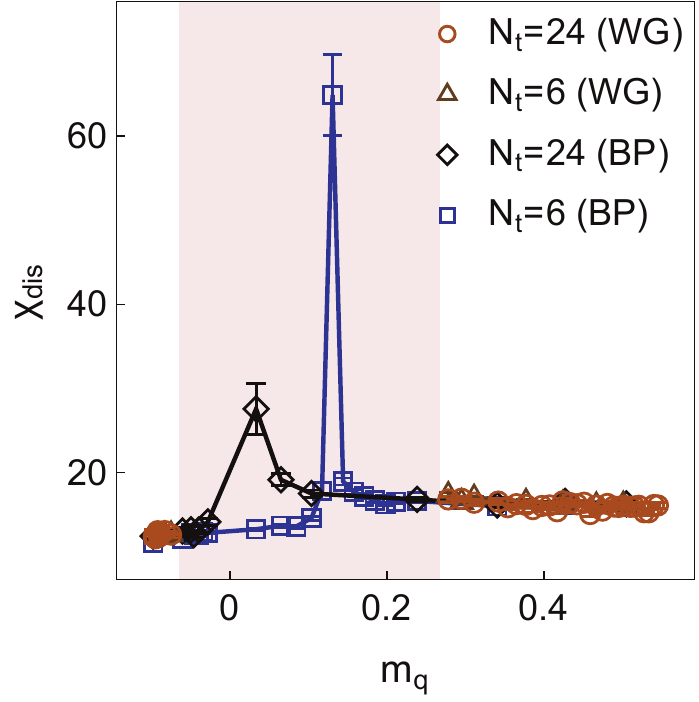}
\end{minipage}\hfill
\begin{minipage}[t]{0.3\linewidth}
\vspace{0pt}
\centering
\includegraphics[height=0.96\linewidth,keepaspectratio,right]{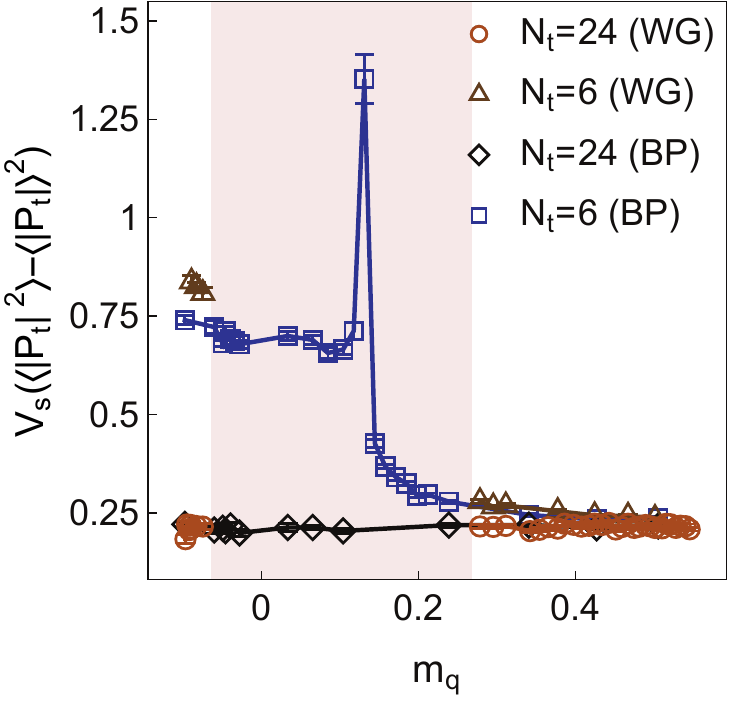}
\end{minipage}\hfill
\begin{minipage}[t]{0.3\linewidth}
\vspace{0pt}
\centering
\includegraphics[height=0.96\linewidth,keepaspectratio,right]{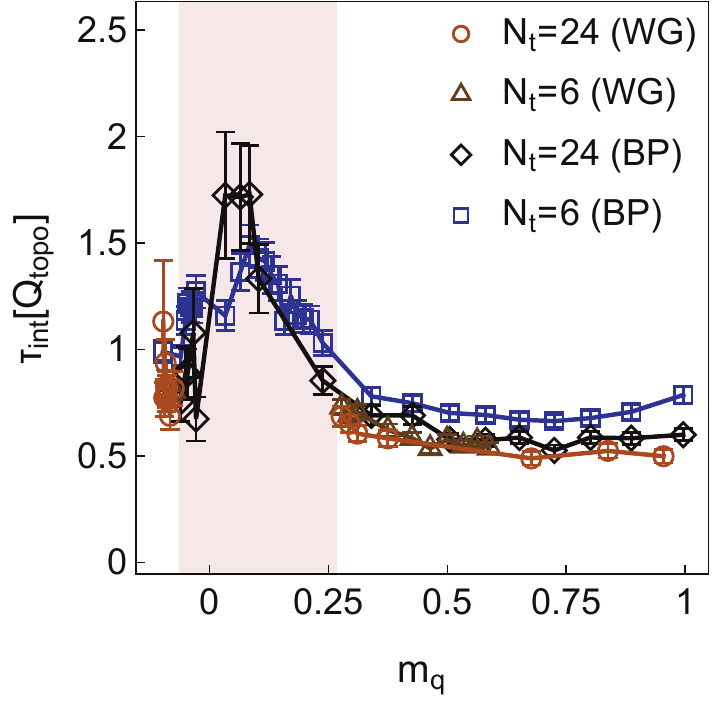}
\end{minipage}
\caption{Simulation results for $\SU{3}$ lattice gauge theory, coupled via 2-step stout smeared links to $N_f=4$ degenerate Wilson clover fermion flavors with hopping parameter $\kappa=0.1358$. The simulations were carried out on lattices of size $12^3\times 6$ (finite temp.) resp. $12^3\times 24$ (zero temp.). As in the previous figures, the orange circles and brown triangles correspond, respectively, to zero and finite temperature results obtained with the Wilson gauge action (WG), Eq.~\eqref{eq:wilsongaugeaction}, and the black diamonds and blue squares to corresponding results obtained with the bulk-preventing action (BP), Eq.~\eqref{eq:bpaction}. Data points are connected by straight lines to guide the eye. The first row shows the PCAC quark mass (top left) and average unsmeared (top center) and smeared (top right) plaquette as functions of $\gamma$ (BP) resp. $\beta-4.167$ (WG). The remaining rows show the quantities as functions of the PCAC quark mass (obtained from the $N_t=24$ simulations). The second row shows the chiral condensate (middle left), the temporal Polyakov loop (middle center), and the topological susceptibility (middle right), and the third row shows the disconnected piece of the chiral susceptibility (bottom left), variance of the temporal Polyakov loop (bottom center), and the integrated auto-correlation time of the topological charge (bottom right). The shaded areas in the different panels mark the PCAC quark mass range that cannot be resolved with the WG action for the given simulation parameters, due to the bulk transition.}
\label{fig:actioncompsu3}
\end{figure*}

As a concrete example for a system where this is the case, we consider $\SU{3}$ lattice gauge theory with the WG action and with $N_f=4$ mass-degenerate, dynamical Wilson-clover fermion flavors, that couple to the gauge field via 2-step stout smeared links. Fig.~\ref{fig:phasediagramsu3} shows a schematic $\of{\beta,\kappa}$-phase diagram for this system. The dashed red line marks the location of the bulk transition, the blue line indicates where the PCAC quark mass, $m_q$, crosses zero, and the dashed black line marks the location of the thermal resp. "confinement/deconfinement" transition if the temporal lattice size is set to $N_t=6$. The label "confinement/deconfinement" is put in quotation marks, because we use the temporal Polyakov loop as approximate order parameter for deconfinement, despite the presence of dynamical fermions~\cite{Detar:1998qa,Karsch:1998hr}. In QCD deconfinement is observed to be accompanied by a chrial transition that occurs at the same temperature; for light fermions, this chiral transition dominates, whereas in the heavy fermion limit, the "confinement/deconfinement" transition of pure gauge theory is approached.

The curves in Fig.~\ref{fig:phasediagramsu3} are based on parameter scans performed with simulations on lattices of spatial size $V_s=N_s^3$ with $N_s=12$. For the $m_q=0$ and the bulk transition lines, the temporal lattice extent was set to $N_t=24$ (approximating zero-temperature), while for the "confinement/deconfinement" transition, the indicated $N_t=6$ was used. The additional dashed lines in different shades of gray are not based on actual simulations; they merely illustrate how the "confinement/deconfinement" transition line is expected to change if $N_t$ is increased (assuming that also $N_s$ is increased accordingly). 

For values of $\kappa$ above $\sim 0.13$ the line where the PCAC quark mass crosses zero coincides with the bulk transition line. Across these coinciding lines the system undergoes a first order transition and PCAC quark mass never passes through the value $m_q=0$ but jumps discontinuously from positive to negative values across the transition line. This is shown in the top left panel of Fig.~\ref{fig:actioncompsu3} where the PCAC quark mass for the WG action (brown circles) is shown as function of $\beta-4.167$ at \changed{$\kappa=0.1358$}. To the left of the bulk transition line, the system is in the unphysical bulk phase, while to the right of the line the system has unphysical negative PCAC quark mass. Thus, the lattice does not describe any continuum-related physics for $\kappa>0.13$. Only for $\kappa<0.13$ there is a range in $\beta$ for which the system is in the continuum phase and the PCAC quark mass is non-negative.

With $N_t=6$, also the "confinement/deconfinement" line in Fig.~\ref{fig:phasediagramsu3} is for \changed{$\kappa<0.142$} on top of the bulk transition line. In the displayed range of $\kappa$ the "confinement/deconfinement" transition separates from the bulk transition line only for \changed{$\kappa>0.142$}, but is then located in the negative PCAC mass region and hence unphysical. 
To extract information about the continuum theory form this lattice system, one would have to increase $N_t$ (and, correspondingly, also $N_s$ to avoid dominance of finite volume effects) so that the "confinement/deconfinement" line fully separates from the bulk transition line.
Thus, we can conclude that it is not possible to reach the light quark confinement (chiral) phase transition with the WG action using $N_t=6$ lattices.
Of course, in the limit $\of{\kappa\to 0}$, where the quark mass grows much larger than the deconfinement energy scale, the "confinement/deconfinement" line is expected to separate from the bulk transition line also for $N_t=6$, as the fermions decouple and the system reduces to pure gauge $\SU{3}$. 

Fig.~\ref{fig:actioncompsu3} contains also results obtained with
the bulk-preventing action \eqref{eq:bpaction}.  In this case the bulk transition is absent and the PCAC quark mass approaches $m_q=0$ continuously. The small gap in the data around $m_q=0$ is due to the slowing down caused by the appearance of zero eigenmodes of the Wilson-Dirac operator when $m_q\to 0$. This could be avoided by e.g. using Schr\"odinger functional boundary conditions, which remove zero modes.

In the first two panels of the second and third row of Fig.~\ref{fig:actioncompsu3} the chiral condensate (2nd row, left) and disconnected chiral susceptibility (last row, left), as well as the temporal Polyakov loop (2nd row, center) and corresponding variance (last row, center) are plotted as functions of the PCAC quark mass, $m_q$. While deep in the strong-coupling and deep in the negative mass phase the results obtained with the WG action agree with those obtained using the BP action, the discontinuity in $m_q$ with the WG action (marked by the shaded areas) implies that the WG action cannot be used to study the transition region. On the other hand, with the BP action there is no discontinuity in $m_q$ and no bulk transition, and the behavior of the chiral condensate and the Polyakov loop at the finite-temperature phase transition are resolved.

The remaining two panels of Fig.~\ref{fig:actioncompsu3}, which show the topological susceptibility (2nd row, right) and integrated auto-correlation time for the topological charge itself (last row, right), indicate that also when coupled to fermions, fluctuations of the gauge-topology are not hindered by the use of the BP gauge action from \eqref{eq:bpaction} and HMC updates. For the values of $m_q$ which are accessible with both actions, both actions yield the same results for the topological susceptibility. The slightly higher integrated auto-correlation time for the topological charge with the BP action at $N_t=6$ is mostly due to a different tuning of the acceptance rates for the HMC trajectories.   
\FloatBarrier

\section{Conclusions}

We have identified a mechanism which appears to be relevant for the formation of unphysical "bulk" configurations and the corresponding occurrence of a "bulk transition" in simulations of lattice $\SU{N}$ gauge theories using Wilson's plaquette gauge action. We proposed a one-parameter family of alternative gauge actions, which possess the same continuum limit as the Wilson plaquette gauge action but which, when used in combination with an HMC update algorithm, prevent bulk-configurations from being created. 

We tested our bulk-preventing simulation framework for pure gauge $\SU{2}$, pure gauge $\SU{5}$, and for $\SU{3}$ with $N_f=4$ mass-degenerate Wilson-clover fermion flavors with hopping parameter \changed{$\kappa=0.1358$}, and which are coupled to the gauge field via 2-step stout smeared link variables. We found that in all three cases, the bulk-preventing action \eqref{eq:bpaction} with $n=2$ removes the bulk transition and reproduces at sufficiently weak coupling the same results as the Wilson plaquette action. 

In the case of the fermionic $\SU{3}$ theory, the Wilson gauge action could not be used to study the physical finite temperature phase transition on $N_t=6$ lattices at small quark masses.  This is due to the fact that the bulk transition prevents the system from simultaneously reaching the physical transition region and small quark masses.
On the other hand, with the bulk-preventing action \eqref{eq:bpaction} the bulk transition is absent and $m_q$ can be made arbitrarily small. 
It is also worth noting that the bulk-preventing actions do not seem to hinder any processes required for topology fluctuations.

\acknowledgements

The authors acknowledge support from the Academy of Finland grants
308791, 319066, and 345070. T.~R. is supported by the Swiss National Science Foundation (SNSF) through the grant no.~TMPFP2\_210064. The authors wish to acknowledge CSC - IT Center for Science, Finland, and the Finnish Computing Competence Infrastructure (FCCI) for computational resources.

\appendix

\changed{
\section{Plaquette wrappings and magnetic monopole creation}\label{asec:monopolesandwrappings}

In this appendix we discuss how in $\Un{1}$ lattice gauge theory the plaquette wrapping types~\ref{en:linkwrap} and~\ref{en:nolinkwrap}, introduced in Sec.~\ref{ssec:motivationinu1}, are related to the creation of magnetic Dirac monopoles as defined in~\cite{DeGrand:1980eq}. The situation with monopoles in $\SU{N}$ lattice gauge theory will be addressed in the second part of this appendix. 

\subsection{Situation in $\Un{1}$}\label{assec:situationinu1}
Dirac monopoles in $\Un{1}$ lattice gauge theory, as described in~\cite{DeGrand:1980eq}, live on the dual lattice and are defined by a non-zero net physical magnetic flux through the boundary of a spatial cube. If this flux is non-zero for a given cube, then this cube contains a Dirac monopole. With our conventions, the corresponding operator can be written as:   
\[
M\of{x}=\frac{1}{2\pi}\epsilon_{4\sigma\mu\nu}\of{\arg\of{U_{\mu\nu}\of{x+\hat{\sigma}}}-\arg\of{U_{\mu\nu}\of{x}}}\ ,\label{eq:magmonopole}
\]
which is the lattice analogue of the continuum expression, $M\of{x}=J_{\of{m}}^{0}$, with
\[
J_{\of{m}}^{\rho}\of{x}=\epsilon^{\rho\sigma\mu\nu}\partial_{\sigma} F_{\mu\nu}\of{x}\ ,
\]
i.e. $M\of{x}=\operatorname{div}\of{B\of{x}}$. If one defines integer plaquette variables $n_{x,\mu\nu}\,\in\,\cof{-2,-1,0,1,2}$, so that
\[
\arg\of{U_{\mu\nu}\of{x}}=\Theta_{x,\mu\nu}-2\,\pi\,n_{x,\mu\nu}\ ,
\]
with $\Theta_{x,\mu\nu}$ from~\eqref{eq:totplaqphaseu1}, then $n_{x,\mu\nu}$ counts the amount of Dirac flux passing through the (oriented) plaquette on which $U_{\mu\nu}\of{x}$ is defined. The magnetic monopole operator~\eqref{eq:magmonopole} can then also be written as~\cite{DeGrand:1980eq}
\[
M\of{x}=\epsilon_{4\sigma\mu\nu}\of{n_{x+\hat{\sigma},\mu\nu}-n_{x,\mu\nu}}\ .\label{eq:magmonopole2}
\]

From \eqref{eq:magmonopole} or \eqref{eq:magmonopole2} we see, that in order to produce or destroy a magnetic monopole, one has to change the net Dirac flux through the boundary of a spatial cube. Considering the plaquette wrapping types \ref{en:linkwrap} and \ref{en:nolinkwrap} introduced in Sec.~\ref{ssec:motivationinu1}, a change of the net Dirac flux through the boundary of a spatial cube can only be achieved with plaquette wrappings of type~\ref{en:nolinkwrap}. Fig.~\ref{fig:monopolesandflux} illustrates how a type~\ref{en:nolinkwrap} plaquette wrapping of a spatial plaquette produces in the two adjacent spatial cubes a pair of magnetic monopoles of opposite charge. The monopoles serve, respectively, as source and sink for the Dirac flux passing through the wrapped plaquette. 

\begin{figure}[!htb]
\centering
\includegraphics[width=0.53\linewidth]{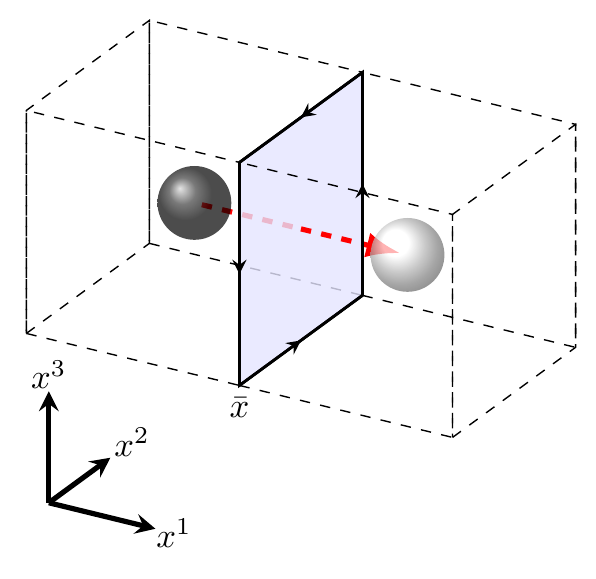}
\caption{\changed{If a single spatial plaquette (blue) picks up a non-zero Dirac flux (red dashed arrow), the two spatial cubes adjacent to the plaquette get, respectively, occupied by a monopole (black sphere) and an anti-monopole (white sphere).}}
\label{fig:monopolesandflux}
\end{figure}

\begin{figure}[!htb]
\centering
\begin{minipage}[t]{0.5\linewidth}
\vspace{0pt}
\centering
\includegraphics[height=1.0\linewidth,keepaspectratio,right]{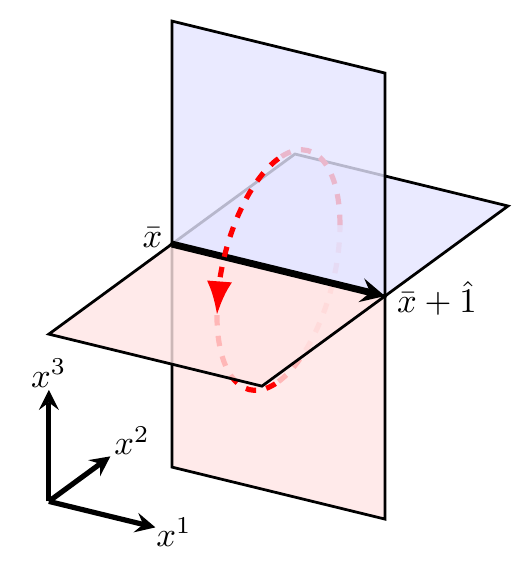}
\end{minipage}\hfill
\begin{minipage}[t]{0.5\linewidth}
\vspace{0pt}
\centering
\includegraphics[height=1.0\linewidth,keepaspectratio,right]{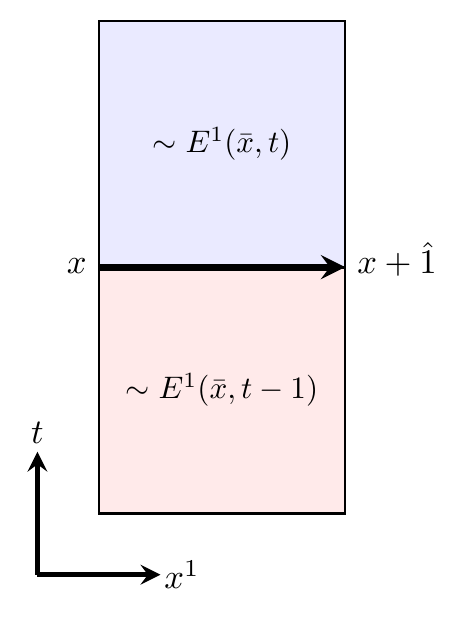}
\end{minipage}
\caption{\changed{If the phase of the displayed spatial link (long black arrow) wraps around the $\loint{-\pi,\pi}$-interval, this implies that the phases of all plaquettes that contain this link also wrap around the $\loint{-\pi,\pi}$-interval. For the spatial plaquettes (left-hand panel) this means that they pick up a Dirac flux passing through them, which is such that a closed magnetic flux loop (red dotted arrow loop) is formed that winds around the phase-wrapped link (long black arrow). For the temporal plaquettes (right-hand panel), the phase-wrapping does not correspond to the production of Dirac flux as temporal plaquettes are associated with the electric and not the magnetic field.}}
\label{fig:closedflux}
\end{figure} 

With plaquette wrappings of type~\ref{en:linkwrap} on the other hand, only closed Dirac strings can be produced and therefore no monopole/anti-monopole pairs are created. The reason for this is, that plaquette wrappings of type~\ref{en:linkwrap} are caused by a link variable wrapping around the $\loint{-\pi,\pi}$-interval. Such a link wrapping will, however, never affect just a single plaquette but all plaquettes that contain the given link. If this link is a spatial one, it is contained in spatial as well as in temporal plaquettes. For the spatial plaquettes the wrapping creates the aforementioned Dirac flux that forms a closed loop winding around the shared link, illustrated in the left-hand panel of Fig.~\ref{fig:closedflux}. 
For the temporal plaquettes, sketched in the right-hand panel of Fig.~\ref{fig:closedflux}, the wrapping does not give rise to the production of Dirac flux as temporal plaquettes are associated to the electric and not the magnetic field. For the same reason, no Dirac flux is produced if the wrapping link is a temporal one, as this would affect only temporal plaquettes.

\subsection{Situation in $\SU{N}$}\label{assec:situationinsun}
In non-abelian $\SU{N}$ gauge theory magnetic monopoles can be defined in different ways~\cite{Goddard:1977da,Brandt:1980em,Weinberg:1983bf,Hollands:1998np} and can be divided into two categories: singular monopoles which require a Dirac string, and non-singular monopoles which possess a Dirac string only in certain gauges. On the lattice singular monopoles can be implemented in terms of net $\mathrm{Z}_N$ Dirac flux through the boundary of spatial cubes~\cite{Ukawa:1979yv,Brower:1981rz,deForcrand:2005zg}, with the Dirac flux, $\sim\exp\of{2\,\pi\,\ii\,n_{x,\mu\nu}/N}$, $n_{x,\mu\nu}\in\cof{0,\ldots,N-1}$, associated with a plaquette variable $U_{\mu\nu}\of{x}$, being defined so that in the factorization
\[
U_{\mu\nu}\of{x}=\e^{2\,\pi\,\ii\,n_{x,\mu\nu}/N}\,\tilde{U}_{\mu\nu}\of{x}\label{eq:znfactor}
\]
one has 
\[
\arg\ssof{\trace\ssof{\tilde{U}_{\mu\nu}\of{x}}}\in\loint{-\pi/N,\pi/N}\ .
\]
The possible values for the first factor on the right-hand side of \eqref{eq:znfactor} correspond to the center of $\SU{N}$: 
\[
\mathrm{Z}\of{\SU{N}}=\cof{\id\,\e^{2\,\pi\,\ii\,n/N}}_{n\in\cof{0,\ldots,N-1}} \cong \mathrm{Z}_{N}\ .\label{eq:suncenter}
\]
As the above discussed Dirac monopoles in $\Un{1}$ lattice gauge theory, also the so defined $\mathrm{Z}_N$ monopoles decouple in the (naive) continuum limit $\of{\beta\to\infty}$ of the $\SU{N}$ lattice gauge theory. Preventing the formation of $\mathrm{Z}_N$ monopoles in $\SU{N}$ lattice gauge theory (with WG action) has for $N=2$ been reported to push the bulk phase to stronger coupling~\cite{Brower:1981rz}. Preventing not only $\mathrm{Z}_N$ monopoles but also the formation of $\mathrm{Z}_N$ Dirac strings from being formed removes the bulk phase completely~\cite{Brower:1981rz,Brower:1982ak,Bornyakov:1991gq}. The continuum limit of the model remains, however, unaffected by these modifications~\cite{Fingberg:1994ut}.   

To make contact with our discussion in Sec.~\ref{ssec:situationinsun} about the relation between the production of bulk configurations in $\SU{N}$ lattice gauge theory  and having $\SU{N}$ plaquettes continuously crossing the cut locus of $\id$ on $\SU{N}$, we note that the non-trivial $\SU{N}$ center elements listed in \eqref{eq:suncenter} are all located beyond this cut locus for $N>2$ (resp. on the cut locus for $N=2$). Therefore, by preventing the continuous cut locus crossing of $\SU{N}$ plaquettes, the production of $\mathrm{Z}_{N}$ Dirac flux though individual plaquettes is not possible, which should also prevent the production of $\mathrm{Z}_{N}$ monopole/anti-monopole pairs. This is the analogue of the type~\ref{en:nolinkwrap} plaquette wrapping from Sec.~\ref{ssec:motivationinu1}, which is responsible for the monopole/anti-monopole pair production in $\Un{1}$ lattice gauge theory, as described in Fig.~\ref{fig:monopolesandflux}.

We note that this does not imply that the use of our bulk-preventing actions~\eqref{eq:bpaction} in combination with HMC updates would prevent the production of $\mathrm{Z}_N$ monopoles in general; only the production of cutoff scale monopoles is exlucded. Extended $\mathrm{Z}_N$ monpoles with "thick" strings~\cite{Mack:1978rq,Mack:1978kr,Brower:1981um} attached should still be possible.

Also, the production of non-singular $\SU{N}$ monopoles~\cite{tHooft:1974kcl,Polyakov:1974ek} is not prevented by our bulk-preventing setup. Non-singular monopoles could on the lattice be located by gauge fixing the system to maximal abelian gauge (MAG)~\cite{Bonati:2013bga} and then performing abelian projection (AP) to the Cartan subgroup $\Un{1}^{\of{N-1}}$~\cite{tHooft:1981bkw}. In each of the $\of{N-1}$ $\Un{1}$ factors of $\Un{1}^{\of{N-1}}$ one can then look for Dirac monopoles with the method from~\cite{DeGrand:1980eq}, discussed in the first part of this appendix. It is only in this maximum abelian gauge that non-singular monopoles show up as Dirac monopoles of the projected $\Un{1}^{\of{N-1}}$ subgroup; fixing to another gauge before projecting to $\Un{1}^{\of{N-1}}$ results in fewer monopoles being found~\cite{Bonati:2010bb} in the $\Un{1}$ subgroup factors. We have not tried to use the MAG-AP method to monitor the monopole densities directly in our HMC simulations with the BP actions from~\eqref{eq:bpaction}. However, there is no reason to expect that the BP setup would prevent continuum-relevant monopoles from being produced. Since finite temperature instantons (calorons) can be described as composite objects, consisting of multiple monopoles~\cite{Lee:1998bb}, the bottom-left panels of Figs.~\ref{fig:actioncompsu2},\ref{fig:actioncompsu5}, which illustrate that topology fluctuations are unhindered by the bulk-prevention, indicate that also the caloron constituent monopoles can fluctuate.}




\section{Computing the gauge force}\label{asec:compgaugeforce}
We note that the plaquette variables satisfy $U^{\dagger}_{\mu\nu}\of{x}=U_{\nu\mu}\of{x}$, so that also the $\Omega_{\mu\nu}\of{x}$ from Eq.~\eqref{eq:omegamat} satisfy $\Omega^{\dagger}_{\mu\nu}\of{x}=\Omega_{\nu\mu}\of{x}$. We can therefore write the bulk-preventing action from Eq.~\eqref{eq:bpaction} as
\[
S_{G,b}\,=\,\frac{\gamma}{n\,N}\sum\limits_{x}\sum\limits_{\mu\neq\nu}\trace\sof{\sof{\Omega^{\dagger}_{\mu\nu}\of{x}\Omega_{\mu\nu}\of{x}}^{-n}-\id}\ .\label{eq:bpactionsymm}
\]
Let us now denote by $\delta^{a}_{y,\rho}$ the variation with respect to the link-variable that lives on the link that points from site $y$ in $\rho$-direction. We then have:
\begin{multline}
\delta^{a}_{y,\rho}\,S_{G,b}\,=\,-\frac{\gamma}{n\,N}\sum\limits_{x}\sum\limits_{\mu\neq\nu}\trace\scof{\sof{\Omega^{\dagger}_{\mu\nu}\of{x}\Omega_{\mu\nu}\of{x}}^{-n}\\
\cdot\sof{\delta^{a}_{y,\rho}\sof{\Omega^{\dagger}_{\mu\nu}\of{x}\Omega_{\mu\nu}\of{x}}^{n}}\\
\cdot\sof{\Omega^{\dagger}_{\mu\nu}\of{x}\Omega_{\mu\nu}\of{x}}^{-n}}\\
=\,-\frac{\gamma}{N}\sum\limits_{x}\sum\limits_{\mu\neq\nu}\trace\scof{\sof{\Omega^{\dagger}_{\mu\nu}\of{x}\Omega_{\mu\nu}\of{x}}^{-\of{n+1}}\\
\cdot\underbrace{\delta^{a}_{y,\rho}\sof{\Omega^{\dagger}_{\mu\nu}\of{x}\Omega_{\mu\nu}\of{x}}}_{\mathclap{\sof{\frac{1}{2}\delta^{a}_{y,\rho}U^{\dagger}_{\mu\nu}\of{x}}\Omega_{\mu\nu}\of{x}+\Omega^{\dagger}_{\mu\nu}\of{x}\sof{\frac{1}{2}\delta^{a}_{y,\rho}U_{\mu\nu}\of{x}}}}}\\
=-\frac{\gamma}{2\,N}\sum\limits_{x}\sum\limits_{\mu\neq\nu}\trace\scof{\sof{\delta^{a}_{y,\rho}U_{\mu\nu}\of{x}}A_{\mu\nu}\of{x}\\
+A^{\dagger}_{\mu\nu}\of{x}\sof{\delta^{a}_{y,\rho}U^{\dagger}_{\mu\nu}\of{x}}}\\
=-\frac{\gamma}{N}\sum\limits_{x}\sum\limits_{\mu\neq\nu}\trace\scof{\sof{\delta^{a}_{y,\rho}U_{\mu\nu}\of{x}}A_{\mu\nu}\of{x}}\ ,\label{eq:gaugeforce1}
\end{multline}
where
\begin{multline}
A_{\mu\nu}\of{x}=\sof{\Omega^{\dagger}_{\mu\nu}\of{x}\Omega_{\mu\nu}\of{x}}^{-\of{n+1}}\Omega^{\dagger}_{\mu\nu}\of{x}\\
=\sof{\Omega^{\dagger}_{\mu\nu}\of{x}\Omega_{\mu\nu}\of{x}}^{-n}\Omega^{-1}_{\mu\nu}\of{x}\ ,
\end{multline}
and we have used that $A^{\dagger}_{\mu\nu}\of{x}=A_{\nu\mu}\of{x}$ as $\Omega_{\mu\nu}\of{x}$ and $\Omega^{\dagger}_{\mu\nu}\of{x}$ commute with each other and with their inverses.
If we now carry out the variation of the plaquette explicitly, we find:
\begin{multline}
\delta^{a}_{y,\rho}U_{\mu\nu}\of{x}=\\
\delta_{x,y}\delta_{\mu\rho}\sof{\delta^{a}U_{\mu}\of{x}}U_{\nu}\of{x+\hat{\mu}}U^{\dagger}_{\mu}\of{x+\hat{\nu}}U^{\dagger}_{\nu}\of{x}\\
+\delta_{x+\hat{\mu},y}\delta_{\nu\rho}U_{\mu}\of{x}\sof{\delta^{a}U_{\nu}\of{x+\hat{\mu}}}U^{\dagger}_{\mu}\of{x+\hat{\nu}}U^{\dagger}_{\nu}\of{x}\\
+\delta_{x+\hat{\nu},y}\delta_{\mu\rho}U_{\mu}\of{x}U_{\nu}\of{x+\hat{\mu}}\sof{\delta^{a}U^{\dagger}_{\mu}\of{x+\hat{\nu}}}U^{\dagger}_{\nu}\of{x}\\
+\delta_{x,y}\delta_{\nu\rho}U_{\mu}\of{x}U_{\nu}\of{x+\hat{\mu}}U^{\dagger}_{\mu}\of{x+\hat{\nu}}\sof{\delta^{a}U^{\dagger}_{\nu}\of{x}}\\
=\ii\,\delta_{x,y}\delta_{\mu\rho}T^{a}U_{\mu}\of{x}U_{\nu}\of{x+\hat{\mu}}U^{\dagger}_{\mu}\of{x+\hat{\nu}}U^{\dagger}_{\nu}\of{x}\\
+\ii\,\delta_{x+\hat{\mu},y}\delta_{\nu\rho}U_{\mu}\of{x}T^{a}U_{\nu}\of{x+\hat{\mu}}U^{\dagger}_{\mu}\of{x+\hat{\nu}}U^{\dagger}_{\nu}\of{x}\\
-\ii\,\delta_{x+\hat{\nu},y}\delta_{\mu\rho}U_{\mu}\of{x}U_{\nu}\of{x+\hat{\mu}}U^{\dagger}_{\mu}\of{x+\hat{\nu}}T^{a}U^{\dagger}_{\nu}\of{x}\\
-\ii\,\delta_{x,y}\delta_{\nu\rho}U_{\mu}\of{x}U_{\nu}\of{x+\hat{\mu}}U^{\dagger}_{\mu}\of{x+\hat{\nu}}U^{\dagger}_{\nu}\of{x} T^{a}\ ,
\end{multline}
where $\cof{T^{a}}_{a=1,\ldots,N^2-1}$ are the hermitian generators of $\SU{N}$, normalized so that $\trace\sof{T^{a}T^{b}}=\delta^{ab}/2$. Plugging this into Eq.~\eqref{eq:gaugeforce1}, we obtain after some manipulations:
\begin{multline}
\delta^{a}_{y,\rho}\,S_{G,b}\,=\,-\frac{2\,\gamma}{N}\sum\limits_{\nu\neq\rho}\repart\sfof{\trace\sof{\ii\,T^{a}U_{\rho\nu}\of{y}A_{\rho\nu}\of{y}}\\
+\trace\sof{\ii\,T^{a}U_{\rho\of{-\nu}}\of{y}A_{\rho\of{-\nu}}\of{y}}
}\ ,\label{eq:gaugeforce2}
\end{multline}
with
\[
U_{\mu\of{-\nu}}\of{x}=U^{\dagger}_{\nu}\of{x-\hat{\nu}}U_{\nu\mu}\of{x-\hat{\nu}}U_{\nu}\of{x-\hat{\nu}}\ 
\]
being the plaquette that starts and ends at site $x$ and is spanned by the $\mu$ and the negative $\nu$ direction, and the corresponding A-matrix,
\[
A_{\mu\of{-\nu}}\of{x}=U^{\dagger}_{\nu}\of{x-\hat{\nu}}A_{\nu\mu}\of{x-\hat{\nu}}U_{\nu}\of{x-\hat{\nu}}\ .
\]

\end{document}